\documentclass[aps,preprint,nofootinbib,preprintnumbers,superscriptaddress]{revtex4}
\usepackage{amssymb}
\usepackage{amsmath}
\usepackage{amsfonts}
\usepackage{graphicx}
\usepackage[hang,tight,raggedright]{subfigure}
\usepackage{multirow}
\usepackage{subfigure}
\usepackage{float}
\usepackage{graphics}
\usepackage{slashed}
\usepackage{color}
\usepackage{bm}
\usepackage{braket}
\usepackage{psfrag}
\usepackage{booktabs}
\usepackage{latexsym}
\allowdisplaybreaks

\def\pslash{p\!\!\!\slash}

\definecolor{darkred}{rgb}{0.7,0.0,0.0}
\definecolor{darkblue}{rgb}{0.0,0.0,0.9}
\definecolor{darkgreen}{rgb}{0.0,0.5,0.0}
\definecolor{brown}{rgb}{0.0,0.0,0.0}

\newcommand{\Dfbd}{\mathord{\buildrel{\lower3pt\hbox{$\scriptscriptstyle\leftrightarrow$}}\over {D}_{\mu}}}

\def\Eslash{E\!\!\!\!\slash}
\newcommand{\be}{\begin{equation}}
\newcommand{\ee}{\end{equation}}
\newcommand{\besp}{\begin{equation}\begin{split}}
\newcommand{\eesp}{\end{split}\end{equation}}

\begin{document}
\preprint{CTPU-17-34}	
\title{Probing the baryogenesis and dark matter relaxed in phase transition by gravitational waves and colliders}
	
\author{Fa Peng Huang}
\email{huangfp@ibs.re.kr}
\affiliation{Center for Theoretical Physics of the Universe, Institute for Basic Science (IBS),
		Daejeon 34051, Korea}
	
\author{Chong Sheng Li}
\affiliation{Department of Physics and State Key Laboratory of Nuclear Physics and Technology,\\
		Peking University, Beijing 100871, China}
\affiliation{Center for High Energy Physics, Peking University, Beijing 100871, China}

\begin{abstract}
The cosmological phase transition with Q-balls production mechanism can explain the baryogenesis and dark matter simultaneously,
where constraints on dark matter masses and reverse dilution are significantly relaxed.
We study how to probe this scenario by collider signals at QCD next-to-leading order and gravitational wave signals.
\end{abstract}
\maketitle
	
\section{introduction}
A longstanding issue in cosmology and particle physics is understanding the
nature of dark matter (DM) and the origin of baryon asymmetry of the universe
(BAU), which is quantified by $\eta_B \equiv n_B/s\sim 10 ^{-10}$~\cite{Ade:2013zuv,Olive:2016xmw} from
the experiment's data of the big bang nucleosynthesis.
To produce the observed BAU, the well-known Sakharov conditions for successful baryogenesis (baryon number violation, C and CP violation,
and departure from equilibrium dynamics or CPT violation)~\cite{Sakharov:1967dj} are necessary.
There are various baryogenesis mechanisms~\cite{Dine:2003ax} to provide these three conditions,
such as  grand unified theory baryogenesis, Affleck-Dine baryogenesis, electroweak baryogenesis, leptogenesis and so on.
On the other hand, the absence of DM signal in DM direct detection experiments
may give us a hint that there may be some new approaches to probe the DM, such as gravitational waves (GWs) experiments.
In this work, we try to use the GWs and collider signals
to probe the baryogenesis mechanism, which can explain the BAU and DM simultaneously
and associates a strong first-order phase transition (FOPT)~\cite{Krylov:2013qe} at several TeV scale with Q-balls~\cite{Friedberg:1976az,Friedberg:1976me, Friedberg:1976ay} generation to relax the constraints.
Most of the mechanisms to simultaneously solve the baryogenesis and DM puzzles usually have two strong constraints, which are
systemically discussed in Ref.~\cite{Shuve:2017jgj}.
One constraint is that the DM mass is usually several GeV,
and the other constraint is that in the most cases the baryon asymmetry produced by heavy particles decays in the early universe
should not be destroyed by inverse washout processes.
In order to guarantee the efficiency production of the baryon asymmetry from heavy particle decay,
we need to tune the reheating temperature carefully.
A strong FOPT with Q-balls production can be used to relax the two constraints~\cite{Krylov:2013qe}, since
the mass of the DM candidate can be larger than TeV in the symmetry broken phase due to the strong FOPT~\cite{Shuve:2017jgj}
and the strong FOPT induced Q-balls can quickly packet the DM candidates into the Q-balls to greatly reduce the inverse dilution~\cite{Krylov:2013qe}.
In this phase transition scenario, phase transition GWs are be produced during the strong FOPT,
which may provide a new approach to probe the new physics beyond the standard model (SM) after the discovery of GWs by aLIGO~\cite{Abbott:2016blz}.
Constraints from the current LHC data~\cite{Demidov:2014mda}, and predictions at future LHC are also
studied in detail in this paper.
The signals and backgrounds with QCD next-to-leading order (NLO) accuracy will also be investigated in this work.
GWs signals and collider signals will provide a realistic and complementary test on this scenario.
	
In Sec.~\ref{sec:EFT}, we describe the effective Lagrangian in the framework of effective field theory (EFT),
and show that the effective operators can explain the BAU and the DM simultaneously.
In Sec.~\ref{sec:gw}, we discuss concrete realization of the FOPT relaxed mechanism and
calculate the phase transition GWs signals in the parameter spaces allowed by
the observed BAU and the DM energy density.
In Sec.~\ref{sec:collider}, the constraints and predictions at the LHC are discussed in detail.
Sec.~\ref{sec:sum} contains our final conclusions.

\section{the  simplified scenario for baryogenesis and dark matter}\label{sec:EFT}
\begin{figure}
\begin{center}
			\includegraphics[scale=0.5]{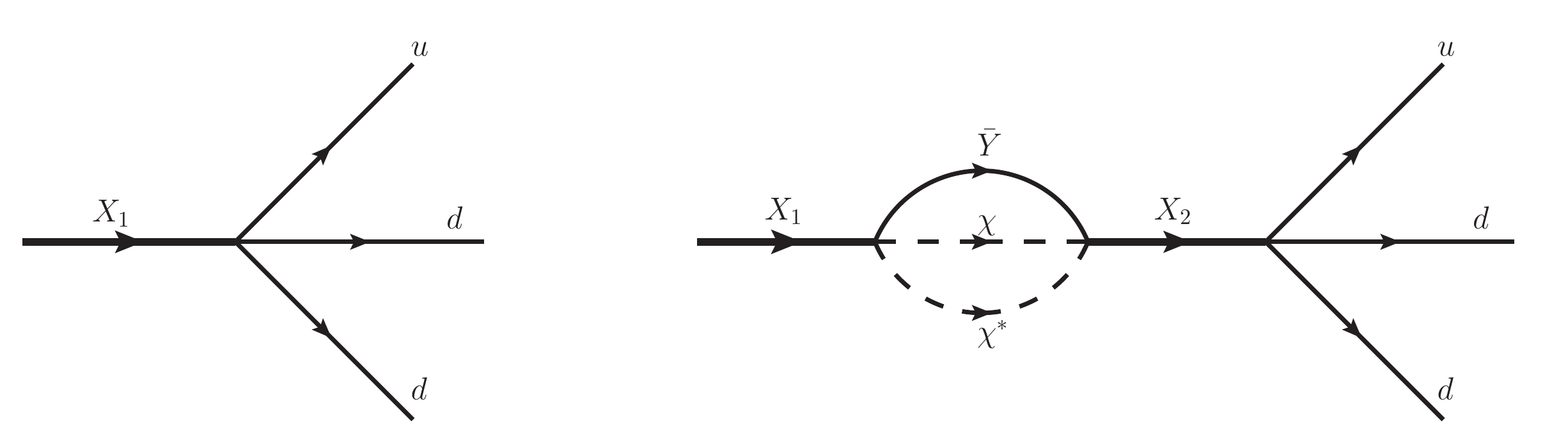}
			\caption{Schematic Feynman diagrams for the production of BAU from the interference effects between tree-level diagram and the two-loop diagram.
			}
			\label{adm}
\end{center}
\end{figure}
In order to explain the baryogenesis and DM simultaneously in this work,
the EFT approach is adopted to provide the model independent
predictions at hadron colliders and GWs detectors.
Firstly, our discussions are based on the following simplified Lagrangian~\cite{Davoudiasl:2010am,Krylov:2013qe},
\begin{eqnarray}\label{effsum}
	\mathcal{L} &=& \frac{1}{2}(\partial_\mu S)^2- U(S)+ (\partial_\mu \chi)^{*}(\partial_\mu \chi)-k_1^2 S^2 \chi^{*} \chi -\sum_{i}\frac{h_i^2}{2} S^2 \phi_i^2\nonumber \\
&+&\sum_{i}\frac{1}{2}(\partial_\mu \phi_i)^2	-\!\!\sum_{a=1,2}\!\!\frac{\lambda^{ijk}_a}{\Lambda^2}\bar{X}_a P_R D_i \bar{U}^{C}_j P_R D_{k} +\frac{\zeta_a}{\Lambda} \bar{X}_a Y^C \chi \chi^* + {\rm H.c.}
\end{eqnarray}
with $U(S)=\lambda_S (S^2-\sigma^2 )^2/4 $.
And $X_a$ represents a heavy Dirac fermionic mediators with several TeV mass, where $a=1,2$ and we assume $m_{X_2} > m_{X_1}$.
The couplings $\lambda^{ijk}_a$ and $\zeta_a$ are complex numbers, which provide the CP violation source.
$X_a$ connects the visible quarks sector and the hidden sector.
$U$ and $D$ represent the up-type quark and down-type quark, respectively.
The dimension-six operator $\frac{\lambda^{ijk}_a}{\Lambda^2}\bar{X}_a P_R D_i \bar{U}^{C}_j P_R D_{k}$ plays important roles
in this scenario and appears in many baryogenesis mechanisms, such as the famous hylogenesis mechanism firstly proposed in Ref.~\cite{Davoudiasl:2010am}.
Collider signals induced by this dimension-six operator have been studied at tree-level using LHC Run-I data in Ref.~\cite{Demidov:2014mda}.
$S$  is a real scalar field, which is the order parameter field for the strong FOPT.
And $\chi$ is a complex field with a global $U(1)$ symmetry.
$\phi_i$ is some unspecified real scalar field, which helps to enhance the strength of the phase transition.
The effective Lagrangian should be realized in some renormalizable UV-completed models, which are left for our future studies.

At the very early universe, the potential $U(S)$ is symmetric due to thermal effects. At this state,
the $S$ field has no vacuum expectation value (VEV), thus the particles $\chi$, $Y$ and $\phi_i$ are massless at tree-level.
At certain time, the non-thermal decays of $X_1$ and $\bar{X}_1$ occur, which will produce baryon asymmetry.
The decay width of the dominant channel for $X_1$ at tree-level is
$X_1\to \bar{Y}\chi \chi^*$ is
\be
\label{A}
\Gamma(X_1\to \bar{Y}\chi \chi^*) = \frac{|\zeta_1|^2\,m_{X_{1}}^3}{1024\,\pi^3 \Lambda^2}\,\,\,\,\,\,\,.
\ee
Another important decay channel is $X_1\to udd$ if only the first generation is considered as an example.
Thus, the corresponding decay width at tree-level can be written as
\be
\label{B}
\Gamma(X_1\to udd)
= \frac{3|\lambda_1|^2\,m_{X_1}^5}{1024\,\pi^3\,\Lambda^4} \,\,\,\,\,\,\,.
\ee

\begin{figure}[h!]
\begin{center}
\includegraphics[scale=0.46]{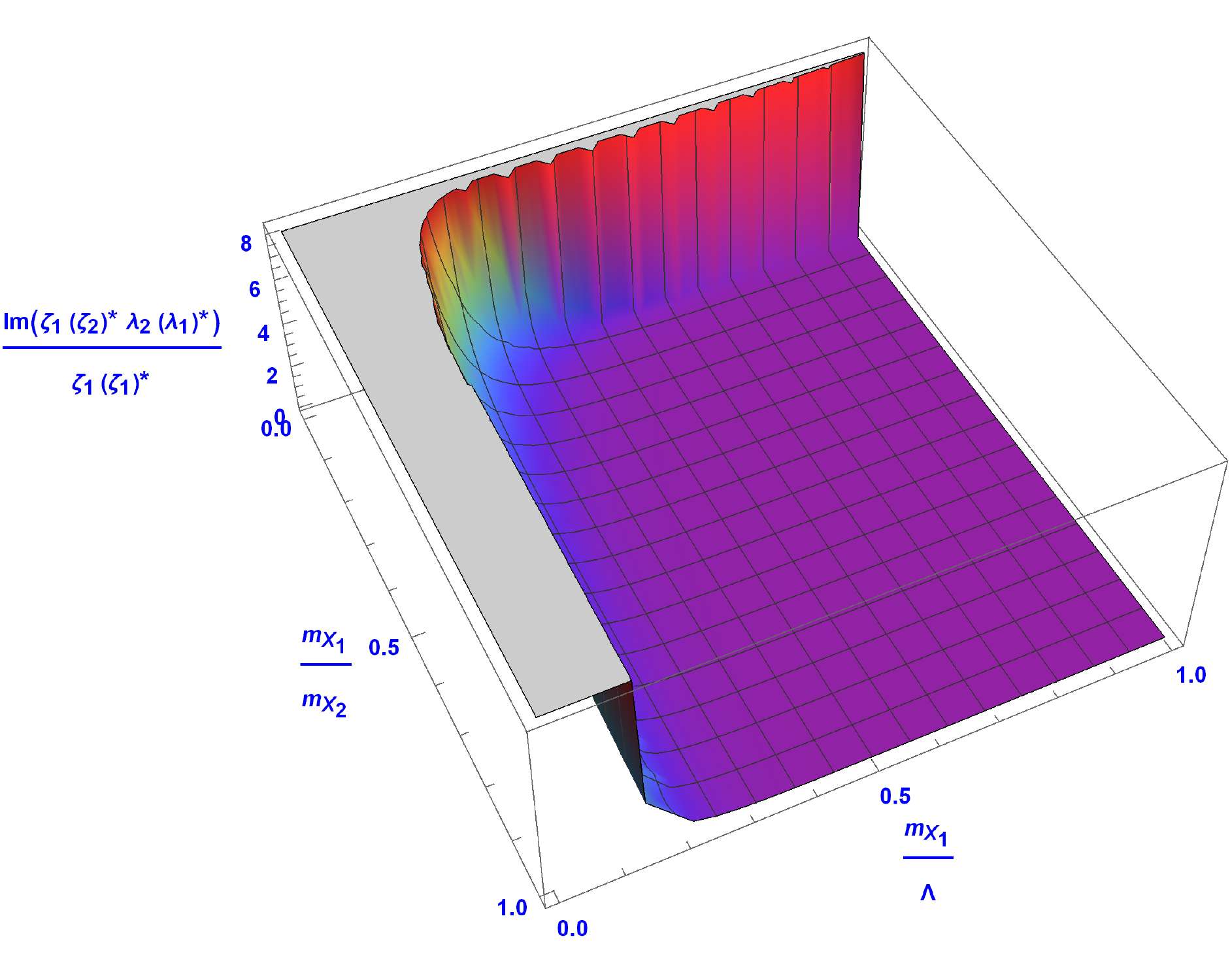}
\caption{Parameter spaces for producing the observed BAU.
The parameter spaces on the colored surface are allowed.}
\label{pa}
\end{center}
\end{figure}

As shown in Fig.~\ref{adm}, the interference effects between the two-loop
diagram and the tree-level diagram produce net baryon asymmetry for per
one $(X_1, \bar{X}_1)$ pair decay, which can be quantified as
\begin{eqnarray}\label{cp_asy}
	\varepsilon &\equiv& \frac{1}{2\Gamma_{X_1}}\left(\Gamma(X_1\to udd) -
	\Gamma(\bar{X}_1\to\bar{u}\bar{d}\bar{d})\right) \nonumber \\
	&\sim & 10^{-5} \times \frac{\rm Im [\lambda^*_1 \lambda_2\zeta_1\zeta^*_2]}{|\zeta_1|^2}
	\frac{m_{X_1}}{m_{X_2}} (\frac{m_{X_1}}{\Lambda})^4 .
\end{eqnarray}	
Essentially, we have $\varepsilon \varpropto \rm Im[\lambda^*_1 \lambda_2\zeta_1\zeta^*_2]$, which represents
the tree-loop interference effects~\cite{Davoudiasl:2010am,Demidov:2014mda}.
Once the asymmetry factor is obtained, the produced BAU can be expressed as
$\eta_B\equiv n_B/s \sim \varepsilon/g_*$.
To satisfy the observed BAU $\eta_B\simeq
10^{-10}$, $\varepsilon \sim 10^{-8}$ is needed for
$g_*\sim 10^2$.
Then, the allowed parameter spaces can be obtained from Eq.~(\ref{cp_asy})
by requiring $\varepsilon \sim 10^{-8}$ for a successful baryogenesis mechanism.
The allowed parameter spaces for producing the observed BAU are shown as the colorful surface in
Fig.~\ref{pa}, where we have $\Lambda>m_{X_2}>m_{X_1}$ for the consistence of the EFT. We
can see that there are no strong constraints on the absolute values of the model parameters
as long as the three ratio values ($\frac{\rm Im [\lambda^*_1 \lambda_2\zeta_1\zeta^*_2]}{|\zeta_1|^2}$,$\frac{m_{X_1}}{m_{X_2}}$,$\frac{m_{X_1}}{\Lambda}$) satisfy a certain relation in Eq.~(\ref{cp_asy}).
	
In this scenario, we have $n_{\chi}=n_{\chi^*} =n_Y=n_B$ after the decay of $X_a$ particles from baryon number conservation.
With the production of BAU, the DM candidate can also be given.  In most mechanisms (we take the hylogenesis mechanism proposed in Ref.~\cite{Davoudiasl:2010am} as a typical example) for explaining DM and BAU simultaneously,
the DM masses should be several GeV~\cite{Davoudiasl:2010am,Demidov:2014mda}.
And the re-scatter effects can wash out the generated baryon asymmetry in the decays
of $X_1, \bar{X}_1$ pair. To suppress this inverse process, additional strong constraints are needed, such
as the requirements of tuning the reheating temperature~\cite{Davoudiasl:2010am,Demidov:2014mda}.
These two constraints can greatly suppress the allowed parameter spaces for successful baryogenesis and DM.
A phase transition mechanism~\cite{Shuve:2017jgj} with Q-balls generation~\cite{Krylov:2013qe} is studied in this work to
avoid these constraints, which are discussed carefully in the following section.
	
\section{Strong First-Order Phase Transition at TeV Scale and Gravitational Waves Signals}\label{sec:gw}
Firstly, we qualitatively describe the scenario that the phase transition with Q-balls generation can relax the above constraints.
After the production of baryon asymmetry from heavy particles decay, we assume that a strong FOPT occurs at several TeV scale by the $S$ field
in Eq.~(\ref{effsum}). Thus, the $S$ field acquires VEV, and the $\chi$ particle obtains large mass.
By assuming that the $\chi$ particle mass in the broken phase is much larger than the critical temperature,
namely, $m_{\chi}=k_1 \sigma \gg T_c$,
$\chi$ particles get trapped in the remnants of the old phase.
Under the assumption $m_{\chi}=k_1 \sigma \gg T_c$, the $\chi$ particle numbers entered into the symmetry breaking phase
are negligibly small due to the exponential suppression $e^{-k_1 \sigma/T_c}$.
And with the bubble expansion, they eventually shrink to very small size objects and become the so-called Q-balls as DM candidates.
As for the particle $Y$, it enters into the symmetry breaking phase and remains massless.
Thus, its contribution to the DM energy density is negligibly and we leave the study on the its roles in the early universe for
our future study.
Particles $\phi_i$ also obtain certain mass $m_{\phi}=h_i \sigma$.
By requiring the condition $ T_c \gtrsim { h_i \sigma,m_S }$, particles $S$ and $\phi_i$ can make efficient thermal
contributions to the strong FOPT.
More explicitly, even when $3 T_c > { h_i \sigma,m_S }$, they can still make some thermal contribution to the FOPT.
Thus, the fundamental requirement for this scenario can be written as
\begin{equation}\label{mr}
k_1 \sigma \gg T_c \gtrsim { h_i \sigma,m_S}   \,\,\,\,.
\end{equation}

\begin{figure}[h!]
\begin{center}
\includegraphics[scale=0.46]{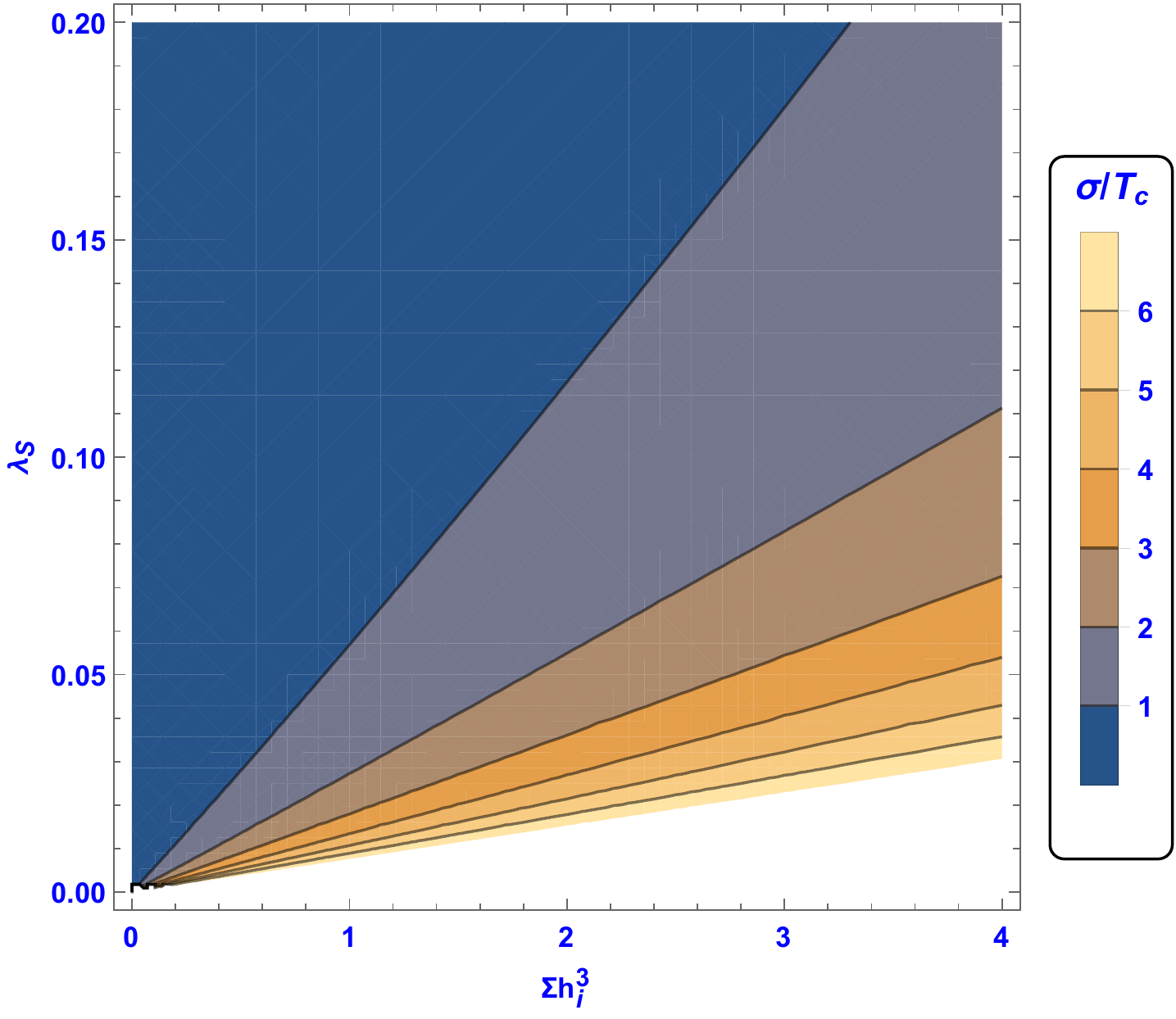}
\caption{Parameter spaces for producing a strong FOPT where the blue region is excluded.}
\label{ws}
\end{center}
\end{figure}

Now, we begin the quantitative investigation from the conditions for a strong FOPT.
From Eq.~(\ref{effsum}), using the standard finite temperature quantum field theory~\cite{Quiros:1999jp}, we can
obtain the following one-loop effective potential at finite temperature
\begin{equation}\label{veff}
V_{\rm eff}(S,T) \approx \frac{\left( - \mu^2_S + c \, T^2 \right) S^2 }{2} - \frac{e \, T (S^2)^{3/2}}{12 \pi}  + \frac{\lambda_S}{4} S^4,
\end{equation}
where $\mu^2_S=\lambda_S \sigma^2$ and $m_S^2=2 \lambda_S \sigma^2$.
The parameter $e$ quantifies the interactions between the $S$ field and the bosons which can make thermal contributions to the phase transition.
Here, the high temperature expansion approximation (namely, the thermal boson function $J_{boson}= -\frac{\pi^4}{45}+\frac{\pi^2}{12}\frac{m^2}{T^2}-\frac{\pi}{6}(\frac{m^2}{T^2})^{\frac{3}{2}}+... $) has been used to obtain the simple results in Eq.~(\ref{veff}). The thermal correction to the coupling $\lambda_S$ is also omitted.
Under these approximations, one can get $e \thicksim \sum_i h_i^3+(3\lambda_S)^{3/2}$ and $c  \thicksim \lambda_S/4+\sum_i h_i^2/12$.
To obtain a strong FOPT, one needs $\sigma(T_c)/T_c \gtrsim 1$ as shown in Fig.~\ref{ws};
namely, one must have
\begin{equation}\label{vtc}
\frac{\sigma(T_c)}{T_c}\thicksim \frac{e}{6\pi \lambda_S} \gtrsim 1,
\end{equation}
where
\begin{equation}\label{tc}
T_c \thicksim \frac{6\pi\mu_S  \sqrt{2\lambda_S}}{\sqrt{-e^2+72 c \pi^2 \lambda_S}}.
\end{equation}
The parameter spaces in the blue region of Fig.~\ref{ws} are excluded by the condition of the strong FOPT.

At the end of the FOPT, the $\chi$ particles
are packed into the so-called Q-balls, which
are compact non-topological soliton objects that exist in some new physics models possessing a global symmetry.
In this work, we consider the Friedberg--Lee--Sirlin type Q-balls~\cite{Friedberg:1976az,Friedberg:1976me, Friedberg:1976ay} and
study whether this type of Q-balls can be give the observed DM density in this scenario.
Here, the Q-balls are generated because the $\chi$ particles just have global
$U(1)$ symmetry\footnote{To avoid the domain wall problem, we assume
the $Z_2$ symmetry is broken.} $\chi \to \mbox{e}^{i\alpha} \chi$.	
The stable Q-balls is a spherical object, where $S=\sigma$ outside the Q-balls and
$S=0$ inside the Q-balls, respectively.
To explain the observed DM energy density, it needs to satisfy the condition
\begin{equation}\label{dmdensi}
\rho_{DM}=m_Q n_Q  \,\,\, ,
\end{equation}
where the current DM mass density $\rho_{DM}\simeq1\times 10^{-6} ~\rm GeV \cdot cm^{-3}$.
To obtain the Q-ball mass $m_Q$, it is necessary to
minimize the following Q-ball energy\footnote{Here, we omit the surface energy of the Q-balls since
the surface energy is much smaller compared to E(R)~\cite{Shuve:2017jgj}.}:
\begin{equation}
E(R) = \frac{\pi Q}{R} + \frac{4\pi}{3} R^3 U_0 \; ,
\label{err}
\end{equation}
where $U_0= \lambda_S \sigma^4/4$.
And by minimizing Eq.~(\ref{err}), the Q-ball mass can be written as~\cite{Krylov:2013qe}
\begin{equation}
	m_Q = \frac{4\sqrt{2} \pi}{3} Q^{3/4} U_0^{1/4}.
\label{rm}
\end{equation}
The stability of the Q-balls needs $m_Q < Q k_1 \sigma $.
Since the non-thermal decays of the heavy particles
give $n_{\chi}=n_{\chi^*} =n_B$, one can see that
\begin{equation}
	\frac{n_Q Q}{s} = 2 \frac{n_B}{s}=2 \eta_B \; ,
	\label{nqq}
\end{equation}
where $\eta_B \sim 10^{-10}$.
From Eqs.~(\ref{dmdensi}) and (\ref{nqq}), we obtain
\begin{equation}
\left. \frac{Q}{m_Q} \right|_{t_0}= \frac{ 2\eta_B s_0}{\rho_{DM}}\,\,\,,
\end{equation}
where the $t_0$  and $s_0$ represent the present value and the current entropy density $s_0\simeq 3000~\rm cm^{-3}$.
Thus, it is necessary to calculate the number density of Q-balls and the typical Q-ball charge
at $T_{\ast}$, which can be obtained by
estimating the volume $V_*$ from which $\chi$ particles are collected
into a single Q-ball.
Based on the fact that the Q-ball volume is the same order as the volume of the
remnant of the symmetry unbroken phase, the radius
$R_*$ of the remnant can be estimated by requiring $R_*^3\Gamma(T)\frac{R_*}{v_b} \sim 1$
for the bubble expansion with velocity $v_b$~\cite{Krylov:2013qe}.
In other words, $R_* \sim (\frac{v_b}{\Gamma(T)})^{\frac{1}{4}} $.
Thus, the Q-ball volume is approximately $V_* = \frac{4\pi}{3} R_*^3$,
and the number density of Q-balls $n_Q=V_*^{-1}$ at $T_{\ast}$ when the phase transition terminates.
From Eq.(\ref{rm}), we can calculate the Q-ball mass.

To clearly see the constraints, we need to know the phase transition dynamics from the previous results.
It is necessary to start with the calculation of
the bubbles nucleation rate per unit volume $\Gamma = \Gamma_0(T) {\mathrm e}^{-S_E(T)}$ and $\Gamma_0(T)\propto T^4$~\cite{Linde:1981zj}.
The Euclidean action $S_E(T)\simeq S_3(T)/T$~\cite{Coleman:1977py,Callan:1977pt}, and then $\Gamma = \Gamma_0 {\mathrm e}^{-S_3 / T}$~\cite{Linde:1981zj}, where
\begin{align}
	S_3(T)
	&= \int d^3x \left[ \frac{1}{2} (\nabla S)^2 + V_{\mathrm{eff}}(S,T) \right]  \,\,\, .
	\label{eq_bounceS3}
\end{align}
From Eq.~(\ref{veff}), the analytic result of $S_3/T$ can be obtained~\cite{Dine:1992wr,Dine:1992vs} as
\begin{equation}\label{s3t}
	\frac{S_3}{T} \approx \frac{13.72 \times 144 \pi^2}{e^2}\left(\frac{-\mu_S^2+c T^2}{2 T^2}\right)^{\frac{3}{2}} f[\frac{-\mu_S^2+c T^2}{2 T^2} \frac{144\pi^2 \lambda_S}{e^2}]
\end{equation}
without assuming the thin wall approximation.
Here, $f(x)=1+\frac{x}{4}[1+\frac{2.4}{1-x}+\frac{0.26}{(1-x)^2}]$.
And the FOPT termination temperature is determined by
\begin{equation}\label{tn}
	S_3(T_{\ast})/T_{\ast} =4\ln (T_{\ast}/100 \mbox{GeV})+137,
\end{equation}
which means the nucleation probability of
one bubble per one horizon volume becomes order 1.
This explains why we can estimate the Q-ball volume $V_*$ when the phase transition terminates in the above discussions.

Combing the above results, the conditions for the observed BAU and DM density give
\begin{equation}\label{bdm}
	\rho_{DM}^4  v_b^{3/4}  =73.5 (2 \eta_B s_0)^3 \lambda_S \sigma^4 \Gamma^{3/4}  \,\,\, \, .
\end{equation}
This equation can give explicit constraints on the model parameters, since $\Gamma(T_{\ast})$ is determined by
the phase transition dynamics which
can be calculated from the original Lagrangian.
As for the bubble wall velocity $v_b$, in principle, it is also depends on the
phase transition dynamics. However, we just take $v_b=0.3$ as the default bubble wall velocity
for simplicity.
For Eq.~(\ref{bdm})
to satisfy the current DM density, the BAU, and
the condition for strong FOPT,
the critical temperature $T_c$ is numerically around several TeV, or
roughly, $1~ \rm TeV < T_c < 20 ~\rm TeV$.
And $k_1$ is about $\mathcal{O}(4)$ from Eqs.(\ref{mr}) and (\ref{bdm}).
We list some benchmark points in Tab.~\ref{benp}.
\begin{table}[h]
\begin{center}
\begin{tabular}{|c|c|c|c|c|c|}
   \hline
   benchmark sets   &$\lambda_S$& $e$    &$c$&$T_c$ [TeV]& $\frac{\sigma}{T_C}$     \\
   \hline
                 I  & 0.008     & 0.754  & 1 & 15.9      &                  5             \\
   \hline
                 II & 0.0016    & 0.151  & 1 & 6.6       &                  5           \\
   \hline
 \end{tabular}
 \end{center}
 \caption{The benchmark sets after considering the combined constraints for producing the observed DM density and BAU with $v_b=0.3$. }\label{benp}
\end{table}

\begin{figure}[h]
\begin{minipage}[t]{0.3\linewidth}
\centering
 \includegraphics[width=1.0\linewidth]{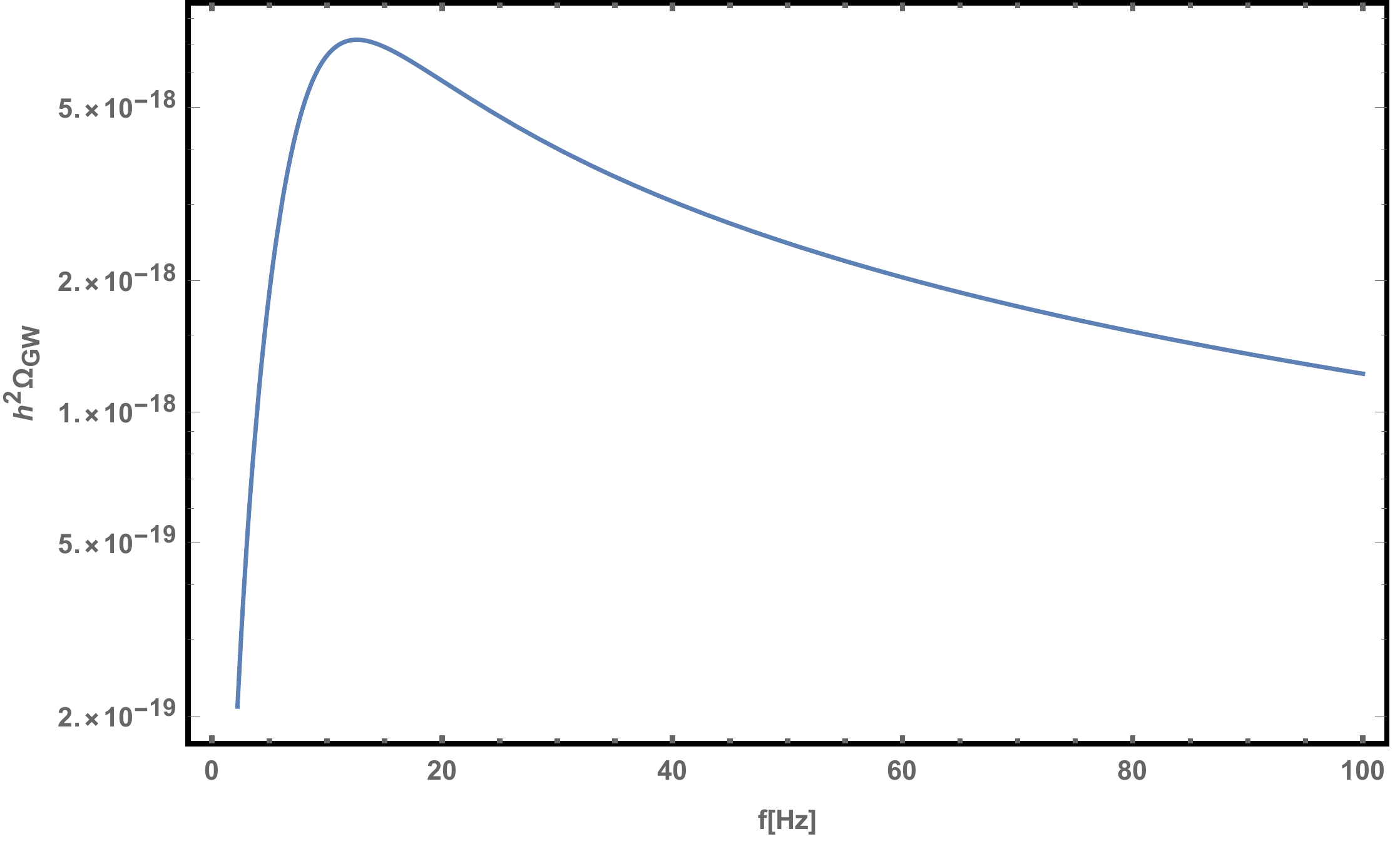}\\
   \text{(a)}
\end{minipage}
\hfill
\begin{minipage}[t]{0.3\linewidth}
\centering
 \includegraphics[width=1.0\linewidth]{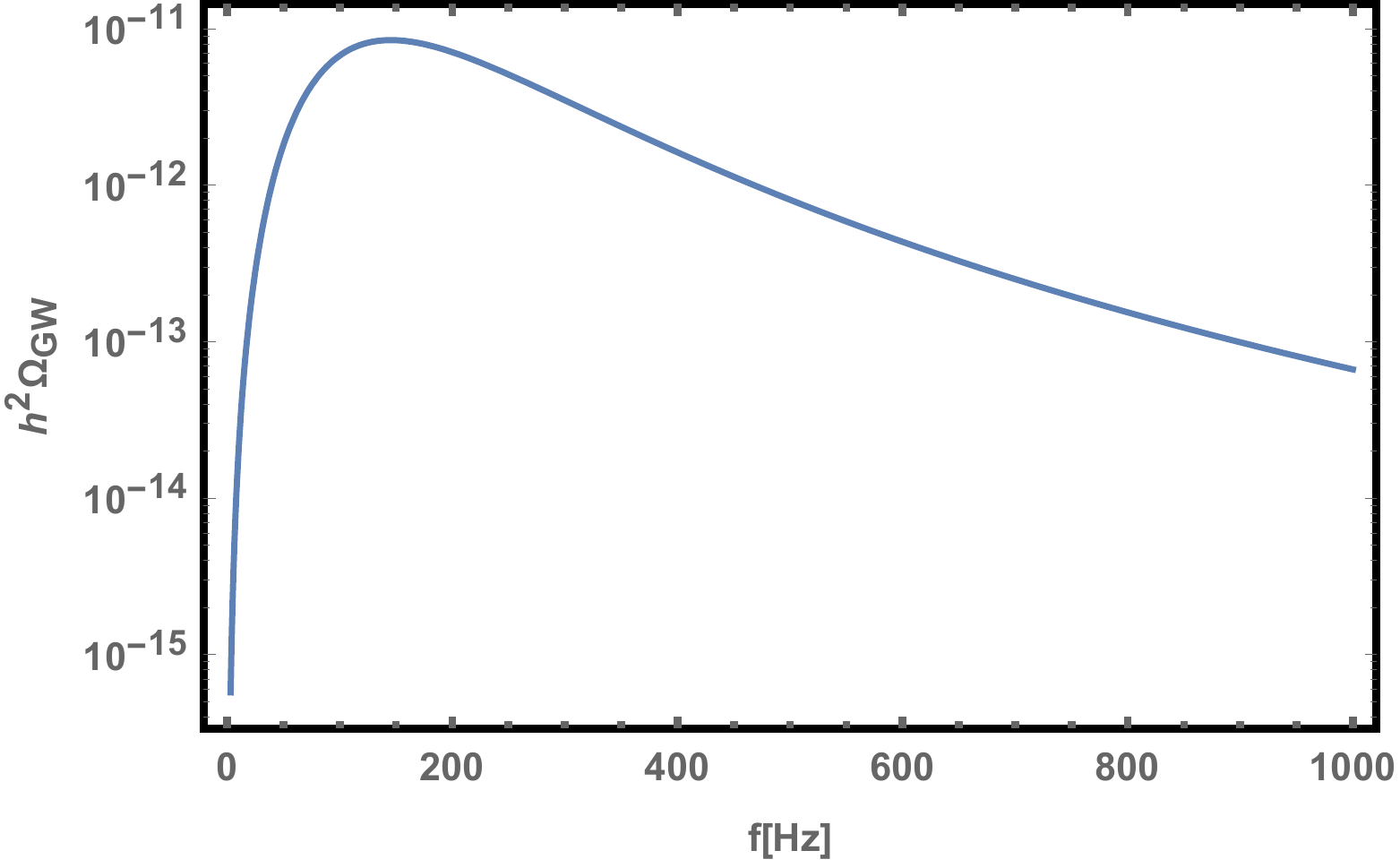}\\
   \text{(b)}
\end{minipage}
\hfill
\begin{minipage}[t]{0.3\linewidth}
\centering
 \includegraphics[width=1.0\linewidth]{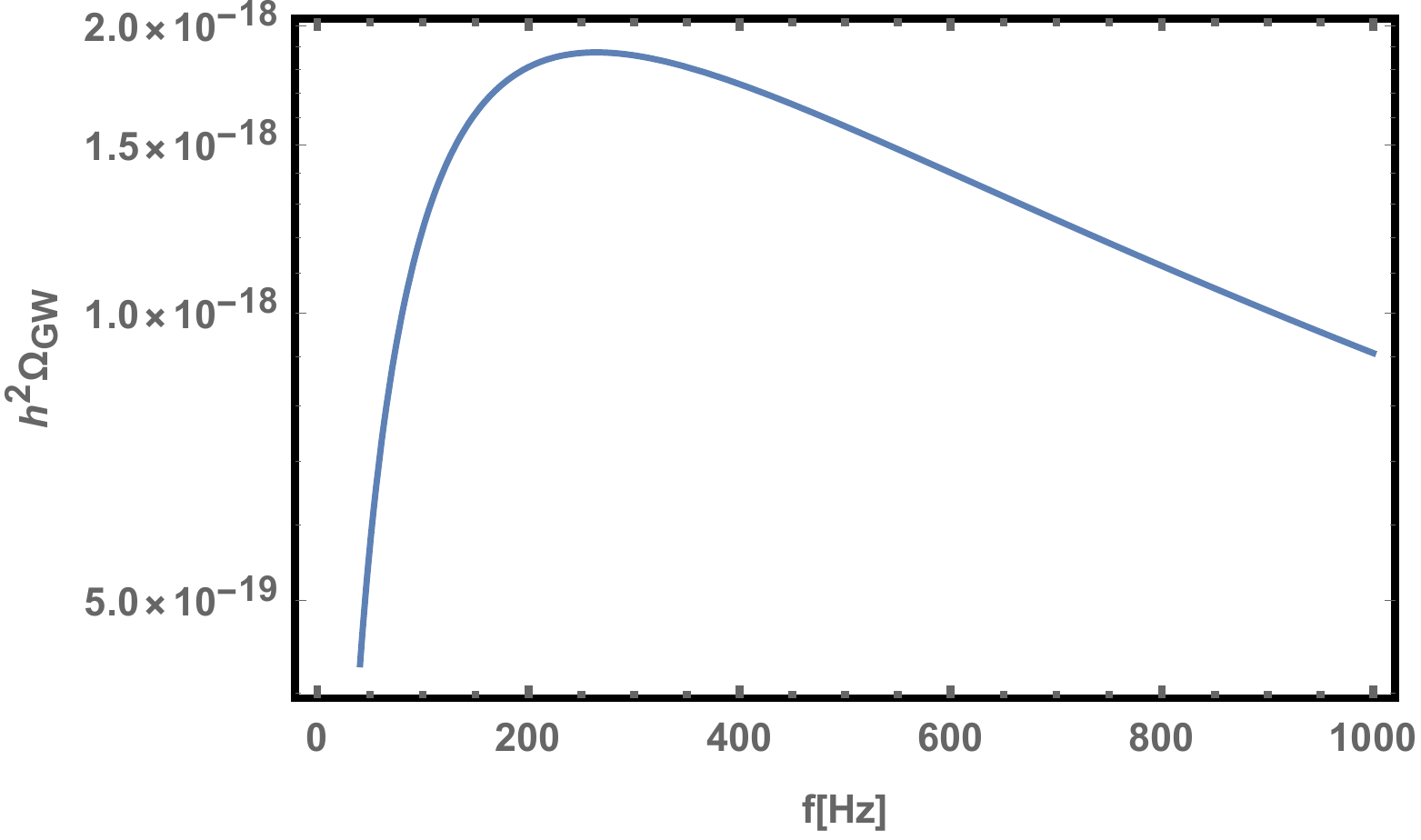}\\
   \text{(c)}
\end{minipage}
\caption{The predicted GWs spectrum for benchmark I with $v_b=0.3$. Figure(a), (b), (c) represents the GWs spectrum from bubble collision, sound waves and turbulence,
respectively.
}\label{gw_b1}
\end{figure}

\begin{figure}[h]
\begin{minipage}[t]{0.3\linewidth}
\centering
 \includegraphics[width=1.0\linewidth]{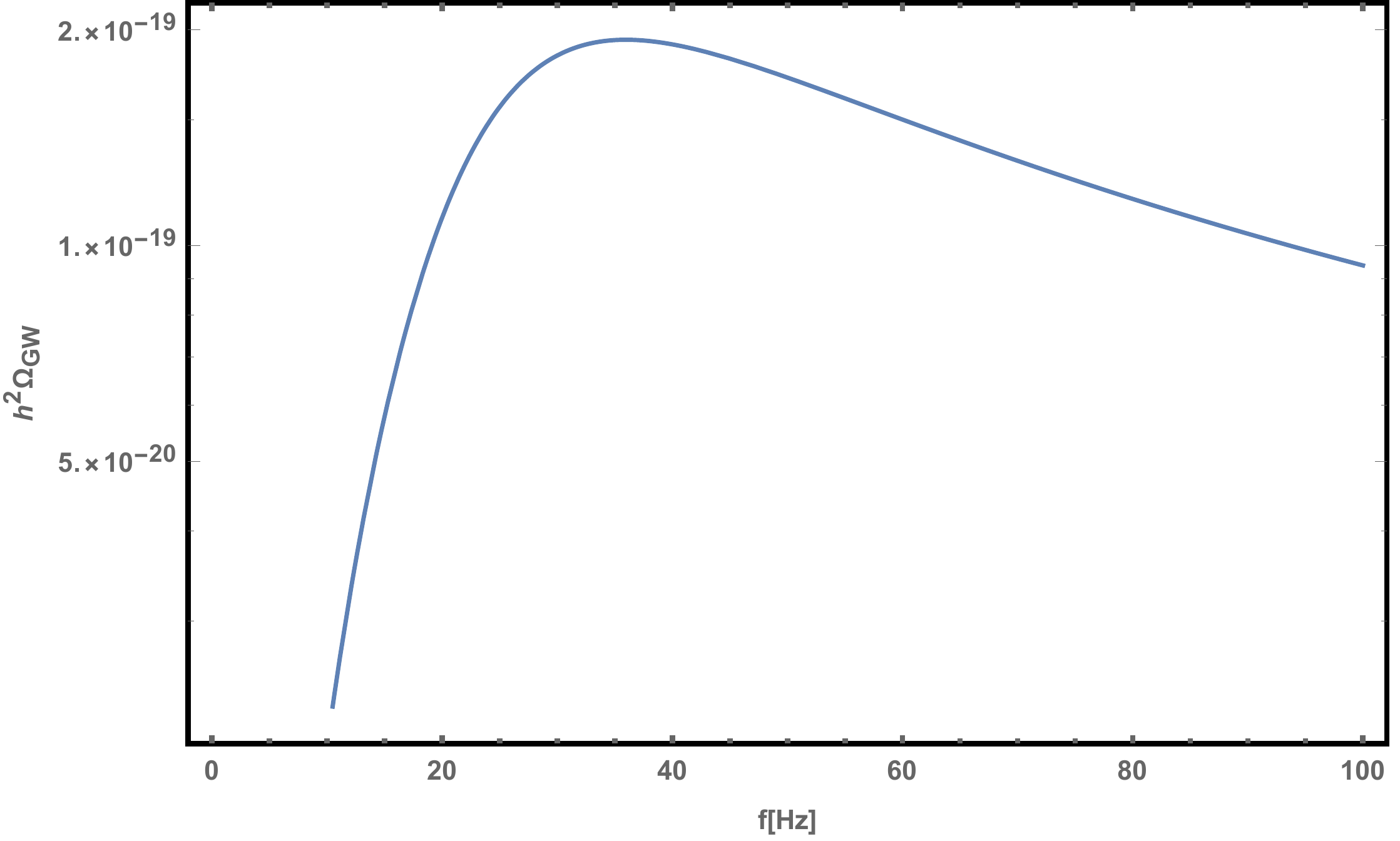}\\
   \text{(a)}
\end{minipage}
\hfill
\begin{minipage}[t]{0.3\linewidth}
\centering
 \includegraphics[width=1.0\linewidth]{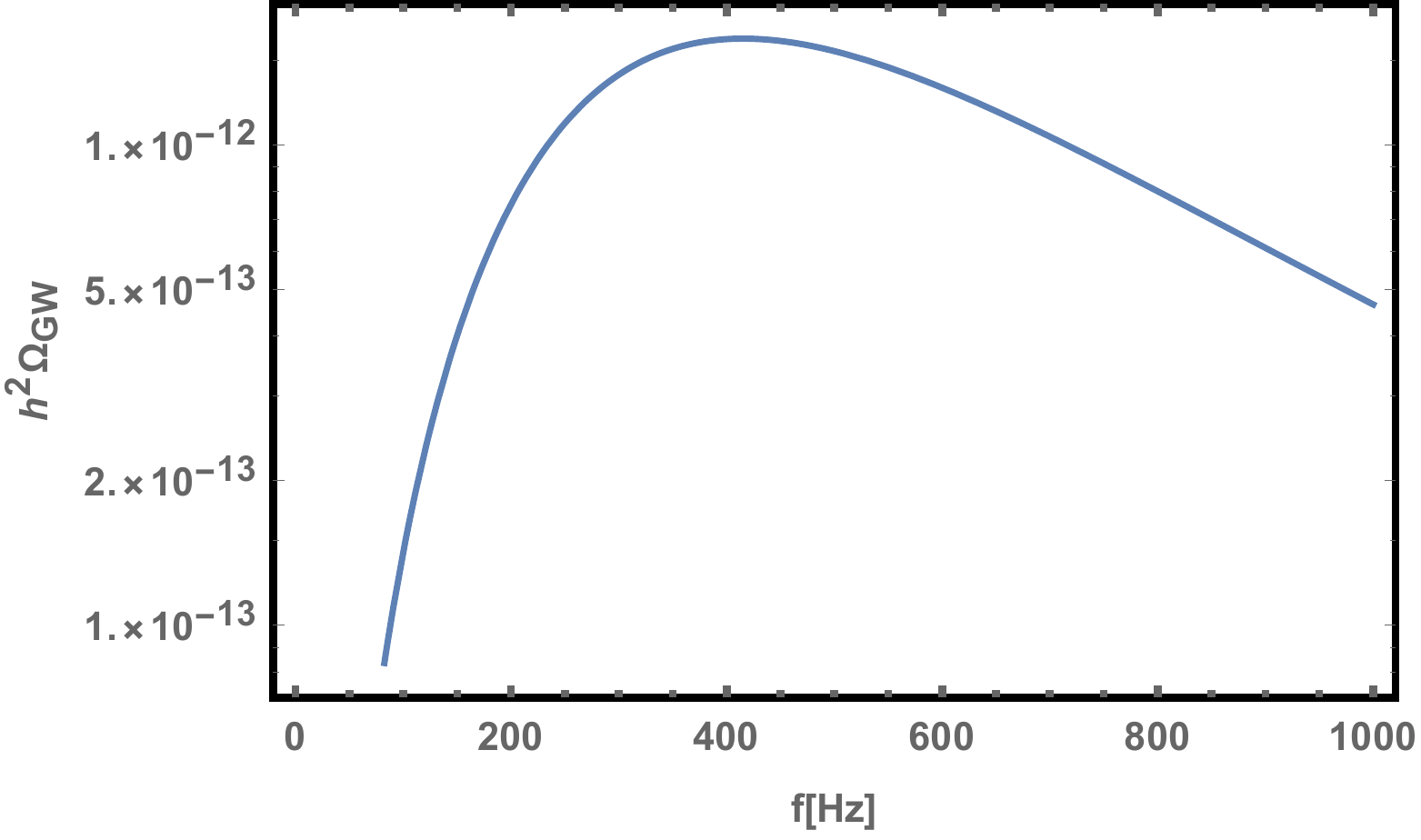}\\
   \text{(b)}
\end{minipage}
\hfill
\begin{minipage}[t]{0.3\linewidth}
\centering
 \includegraphics[width=1.0\linewidth]{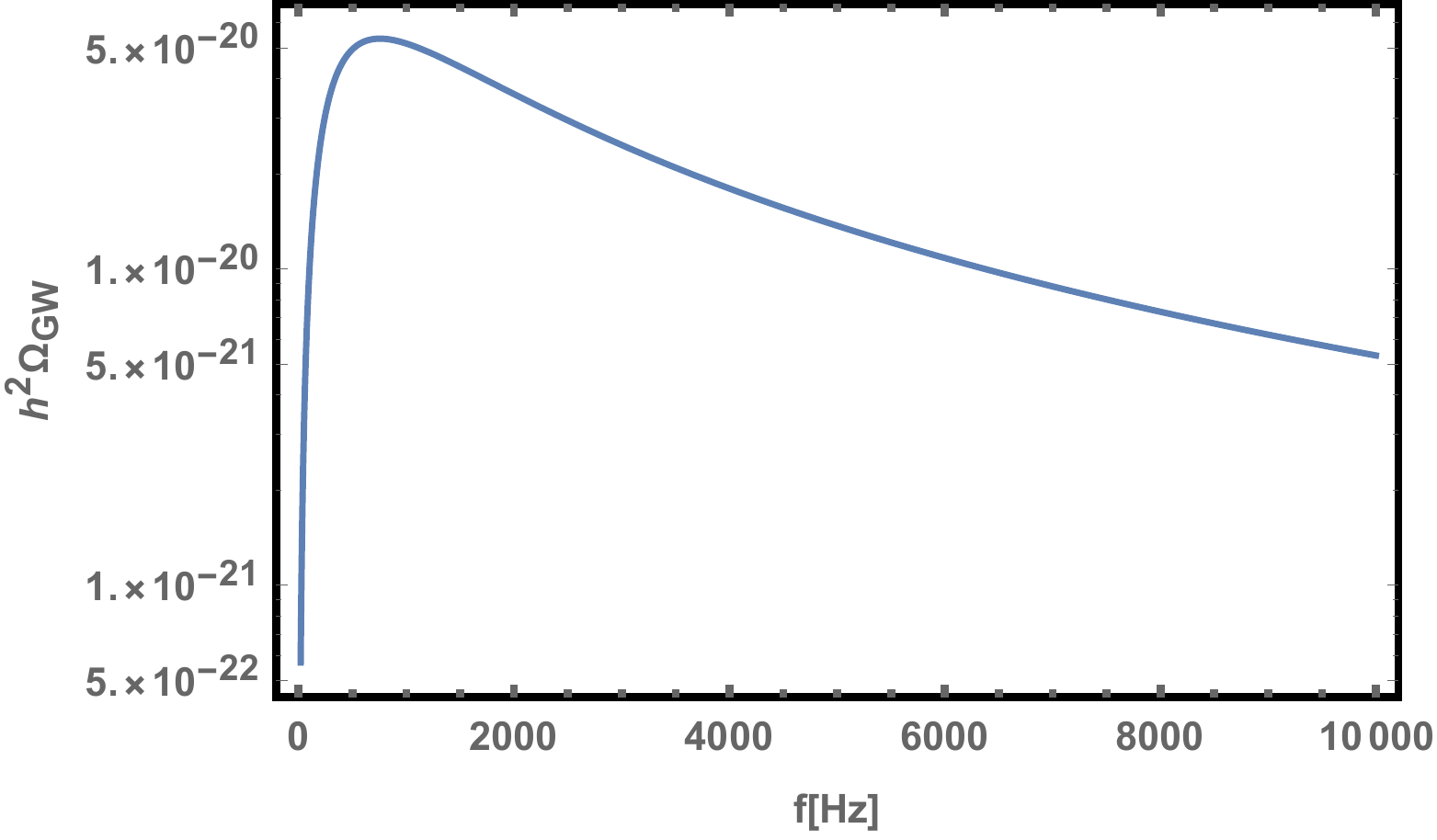}\\
   \text{(c)}
\end{minipage}
\caption{The predicted GWs spectrum for benchmark II with $v_b=0.3$. Figure
(a), (b), (c) represents the GWs spectrum from bubble collision, sound waves and turbulence,
respectively.
}\label{gw_b2}
\end{figure}

Here, there is a strong FOPT at several TeV scale, which 
produces sizable phase transition GWs.
We consider three phase transition GWs sources:
the well-known bubble collisions~\cite{Kamionkowski:1993fg},
the turbulence in the fluid, where a certain fraction of the bubble walls energy is converted into turbulence~\cite{Kosowsky:2001xp, Caprini:2009yp},
and the new source of sound waves~\cite{Hindmarsh:2013xza}.
There are usually four parameters which determine the phase transition GWs spectrum,
namely, $v_b$, $\lambda_i$, $\alpha$ and $\frac{\beta}{H_*}$.
The bubble wall velocity $v_b$ and the energy efficiency factor $\lambda_i$ (i=co, tu, sw) are not easy to
be obtained directly from the Lagrangian, and we just choose some default value or formulae in this work.
The parameters $\alpha$ and $\frac{\beta}{H_*}$ can be directly calculated from the above formulae.
Firstly, the parameter $\alpha \equiv \frac{\epsilon(T_{\ast})}{\rho_{\rm rad}(T_{\ast})}$ is defined at the temperature
$T_{\ast}$ by Eq.(\ref{tn}), wherein $\rho_{\rm rad}(T_{\ast})=\frac{\pi^2}{30} g_{*}(T)T^4$ is the plasma thermal energy density
and $\epsilon(T_{\ast}) = [T \frac{dV_{\rm eff}^{\rm min}}{dT} -V_{\rm eff}^{\rm min}(T) ]|_{T=T_{\ast}}$ is the false vacuum energy density.
A larger $\alpha$ means a stronger FOPT since the strength of the FOPT is measured by the parameter $\alpha$.
Secondly, the parameter $\frac{\beta}{H_*}$ is defined as $\frac{\beta}{H_*}
\equiv \left. T \frac{d(S_3 / T)}{dT} \right|_{T=T_*}$, and
$\beta^{-1}$ represents the typical time duration of the phase transition.
After the four parameters are obtained, we can directly calculate the GWs spectra from the previous three sources including the red-shift effects $\frac{a_\ast}{a_0}= 1.65 \times 10^{-5} \mbox{Hz}\times\frac{1}{H_{\ast}} \Big( \frac{T_{\ast}}{100 \rm GeV} \Big) \Big( \frac{g^t_\ast}{100} \Big)^{1/6}$.
Thus, the peak frequency at current epoch from the three sources can be written as $f_{\rm i}=f_{\rm i}^\ast a_\ast/a_0$.
For the bubble collision, the corresponding $f_{\rm co}^\ast=0.62\beta/(1.8-0.1v_{b} +v_{b}^{2})$~\cite{Huber:2008hg} and
the phase transition GWs spectrum can be written as~\cite{Huber:2008hg,Jinno:2016vai,Jinno:2017fby,Jinno:2017ixd}
\begin{align}
	\Omega_{\rm co} (f) h^2 \simeq
	&1.67\times 10^{-5} \Big( \frac{H_{\ast}}{\beta} \Big)^2 \Big( \frac{\lambda_{co} \alpha}{1+\alpha} \Big)^2 \Big( \frac{100}{g^t_\ast} \Big)^{\frac{1}{3}} \nonumber \\
	&\times \Big( \frac{0.11v_b^3}{0.42+v_b^3} \Big) \Big[ \frac{3.8(f/f_{\rm co})^{2.8}}{1+2.8(f/f_{\rm co})^{3.8}} \Big] .
\end{align}
For the sound wave, the phase transition GWs spectrum can be expressed
as~\cite{Hindmarsh:2013xza, Caprini:2015zlo}
\begin{align}
	\Omega_{\rm sw} (f) h^2\simeq \nonumber
	& 2.65\times 10^{-6}\Big(\frac{H_{\ast}}{\beta}\Big) \Big(\frac{\lambda_{sw} \alpha}{1+\alpha}\Big)^2
	\Big(\frac{100}{g^t_\ast}\Big)^{\frac{1}{3}}v_b\\
	&\times\Big[\frac{7(f/f_{\rm sw})^{6/7}}{4+3(f/f_{\rm sw})^2}\Big]^{7/2}\nonumber
\end{align}
with $f_{\rm sw}^\ast=2\beta/({\sqrt{3}v_b})$ at $T_{\ast}$~\cite{Hindmarsh:2013xza, Caprini:2015zlo},
and for relativistic bubbles~\cite{Espinosa:2010hh} $\lambda_{sw}\simeq \alpha \left(0.73+0.083\sqrt{\alpha}+\alpha\right)^{-1}$.
For the turbulence,  the peak frequency at $T_{\ast}$ is about $f_{\rm tu}^\ast=1.75\beta/v_b$~\cite{Caprini:2015zlo},
and the GWs spectrum is formulated by \cite{Caprini:2009yp, Binetruy:2012ze}
\begin{align}
	\Omega_{\rm tu} (f) h^2\simeq \nonumber
	& 3.35\times 10^{-4}\Big(\frac{H_{\ast}}{\beta}\Big) \Big(\frac{\lambda_{\rm tu} \alpha}{1+\alpha}\Big)^{3/2}
	\Big(\frac{100}{g^t_\ast}\Big)^{\frac{1}{3}}v_b\\
	&\times\frac{(f/f_{\rm tu})^3}{(1+f/f_{\rm tu})^{11/3}(1+8\pi fa_0/(a_\ast H_\ast))}. \nonumber
\end{align}

In Fig.~\ref{gw_b1} and Fig.~\ref{gw_b2}, we show the GWs spectra for the benchmark sets I and II, respectively.
And in each figure, (a),(b),(c) represents the GWs spectrum from bubble collision, sound waves and turbulence,
respectively. We can see that the peak frequency ranges from 20 Hz to several hundred Hz. The large peak frequency
comes from high critical temperature and large $\beta$ which can be seen from the above GWs formulae.
The peak frequencies are just within the region of aLIGO, but the amplitude of the signal is too weak to
be detected by current aLIGO~\cite{Huang:2016odd,Dev:2016feu,Huang:2017laj,Huang:2017rzf}. Future aLIGO-like GWs experiments with even higher precision may help
to probe this type of GWs signals.
%
	
\section{Collider Phenomenology}\label{sec:collider}
Besides the GWs signals discussed above,
we begin to discuss the collider phenomenology at LHC in this section.
From the Lagrangian in Eq.(\ref{effsum}), there are many types of combinations for the
up-type quark and down-type quark, which result in abundant collider phenomenology at the LHC. The interactions between quarks and heavy Dirac fermionic mediator $X_a$ can be described by effective operators
\begin{align}
\label{first-a}
{\cal O}^{dud} & = -\frac{\lambda_a^{dud}}{\Lambda^2}(\bar{X}_a P_R
d)(\bar{u}^{C} P_R d), \\
\label{first-b}
{\cal O}^{dus} & = -\frac{\lambda_a^{dus}}{\Lambda^2}(\bar{X}_a P_R
d)(\bar{u}^{C} P_R s), \\
\label{second-a}
{\cal O}^{dub} & = -\frac{\lambda_a^{dub}}{\Lambda^2}(\bar{X}_a P_R
d)(\bar{u}^{C} P_R b), \\
\label{second-b}
{\cal O}^{dtd} & = -\frac{\lambda_a^{dtd}}{\Lambda^2}(\bar{X}_a P_R
d)(\bar{t}^{C} P_R d).
\end{align}	
At tree level, these operators can result in the following processes
\begin{align}
	u(p_1)\,+\,d(p_2)&\to \bar{d}(p_3)\, +\,X_a(p_4)\,,\\
	u(p_1)\,+\,d(p_2)&\to \bar{s}(p_3) + \,X_a(p_4)\,, \\
	u(p_1)\,+\,d(p_2)&\to \bar{b}(p_3) + \,X_a(p_4)\,, \\
	d(p_1)\,+\,d(p_2)&\to \bar{t}(p_3)\, +\, X_a(p_4)\,,
\end{align}
and their various crossings and charge-conjugated processes. The dominant decay channel of $X_a$ is  $ X_a\to Y \bar{\chi}\chi$, and
$X_a$ behaves as the missing energy in the detector.
The subdominant process of four jet ($X_a$ can decay to three quarks) is not discussed in this work.
Similar collider signals are discussed using LHC Run-I data in Ref.~\cite{Demidov:2014mda} at tree-level.
So the interactions can be explored by performing mono-jet and mono-top analysis at the LHC.
Because the LHC is a proton-proton collider with high precision, the QCD  NLO predictions for these processes are necessary in order to obtain reliable results. In this section, we calculate the NLO QCD corrections for all of the above processes, and investigate the constraints on the interactions described by Eq.~(\ref{first-a}). For convenience, when there is no special description,
the new physics scale $\Lambda$ is fixed at 5 TeV and the dimensionless coupling $\lambda_{dqq}$ is set as 1.
Here,  the parton distribution function (PDF)  NNPDF30nlo~\cite{Ball:2014uwa} is used and top quark mass is fixed at 173 GeV.
To compare with the parameter spaces allowed by the conditions of successful baryogenesis and DM, we just need to rescale these parameters.
	
\subsection{NLO QCD calculations}
The NLO QCD correction can be expressed as
\begin{equation}
\begin{aligned}
\sigma_{\rm NLO}=&\sigma_R + \sigma_V\\
=&\int d\Gamma_3 |{\cal M}_{2\to3}|^2+\int d\Gamma_2 |{\cal M}_{2\to2}|^2\,,
\end{aligned}
\end{equation}
where $d\Gamma_n$ denotes $n$-body phase space. By two cutoff phase space slicing method \cite{Harris:2001sx}, the real radiation in $\sigma_{R}$ can be divided into soft, collinear and hard regions
\begin{equation}
\begin{aligned}
\sigma_R =\sigma_S + \sigma_C + \sigma_H\,,
\end{aligned}
\end{equation}
where $\sigma_{S,C,H}$ depend on two cutoff parameters  $\delta_s$ and $\delta_c$.
With dimension regularization, the hard contribution is finite, which can be calculated numerically. While $\sigma_{S}$ and $\sigma_{C}$ suffer from soft and collinear divergences, which cancel with the infra-red (IR) singularity in the virtual correction $\sigma_V$. So the sum of all the contributions is IR safe.  Using this approach, we firstly give the analytical results of one-loop virtual correction for mono-jet and mono-top productions in Appendix~\ref{app1}.
	
The numerical results of mono-jet processes induced by ${\cal O}_{dud}$ and ${\cal O}_{dus(b)}$ are listed in Tabs.~\ref{nloresXdud} and \ref{nloresXdus}, respectively.  The events are selected with jets in region $p_{T,j}> 250\,{\rm  GeV}$ and $|\eta_j|<2.4$, which is consistent with the kinematic cuts used in Ref.~\cite{ATLAS:2017dnw}. The factorization and renormalization scales are fixed at 1 TeV. The cross section of the mono-jet process induced by ${\cal O}_{dud}$ is significantly larger than the one of ${\cal O}_{dus}$ because the parton density of $u$ quark is much larger than $s$ quark. The $K$-factor decreases with increasing mass of Dirac fermionic $X_a$ and missing transverse energy cut $\Eslash_T$. The differences of the cross section between the processes induced by ${\cal O}_{dus}$ and ${\cal O}_{dub}$ is very tiny, because $b$ quark mass can be neglected at high energy and the PDFs of strange and bottom quark are much smaller than $u$ and $d$.
\begin{table}
		\begin{center}
			\begin{tabular}{ c | c  c  c||  c  c  c}
				\hline
				\hline
				&  & $\Eslash_T>$700 GeV  & & &   $\Eslash_T>$1 TeV &
				\\
				\hline
				~~	$m_X$[TeV]~~ &~~ $\sigma_{\rm LO}[{\rm fb}]$~~ & ~~$\sigma_{\rm NLO }[{\rm fb}]$~~ & ~~$K$-factor~~&~~$\sigma_{\rm LO}[{\rm fb}]$~~ &~~$\sigma_{\rm NLO}[{\rm fb}]$~~ & ~~$K$-factor~~
				\\
				1.2  & 5.49  & 5.63 & 1.02  & 3.61 & 3.64 & 1.01
				\\
				2    & 2.19    & 2.19    & 1     & 1.51 & 1.49 & 0.99
				\\
				2.8  & 0.766 & 0.748 & 0.98 & 0.54 & 0.52 & 0.96
				\\
				3.6  & 0.241 & 0.23 & 0.076  & 0.17 & 0.16 &0.94
				\\
				4  & 0.13 & 0.123 & 0.95     & 0.091 & 0.085  & 0.93
				\\
				\hline
				\hline
			\end{tabular}
		\end{center}
		\caption{\label{nloresXdud} Fixed order results for mono-jet process induced by ${\cal O}_{dud}$ at the 13 TeV LHC.}
\end{table}
	
	\begin{table}
		\begin{center}
			\begin{tabular}{ c | c  c  c||  c  c  c}
				\hline
				\hline
				&  & $\Eslash_T>$700 GeV  & & &   $\Eslash_T>$1 TeV &
				\\
				\hline
				~~	$m_X$[TeV]~~ &~~ $\sigma_{\rm LO}[{\rm fb}]$~~ & ~~$\sigma_{\rm NLO}[{\rm fb}]$~~ & ~~$K$-factor~~&~~$\sigma_{\rm LO}[{\rm fb}]$~~ &~~$\sigma_{\rm NLO}[{\rm fb}]$~~ & ~~$K$-factor~~
				\\
				1.2 & 1.11     & 1.13   & 1.02    & 0.713 & 0.713 & 1.00
				\\
				2   & 0.463  & 0.461 &0.996   & 0.318 & 0.312 & 0.98
				\\
				2.8 & 0.174  & 0.170 & 0.974  & 0.124 & 0.118 & 0.957
				\\
				3.6 & 0.06   & 0.057 & 0.953 & 0.043 & 0.04 & 0.938
				\\
				4    & 0.034 & 0.032 & 0.944 & 0.0245 & 0.0228 & 0.93
				\\
				\hline
				\hline
			\end{tabular}
		\end{center}
		\caption{\label{nloresXdus}Fixed order results for mono-jet process induced by ${\cal O}_{dus(b)}$  at the 13 TeV LHC. }
	\end{table}
	
\begin{figure}
		\begin{center}
			\includegraphics[scale=0.42]{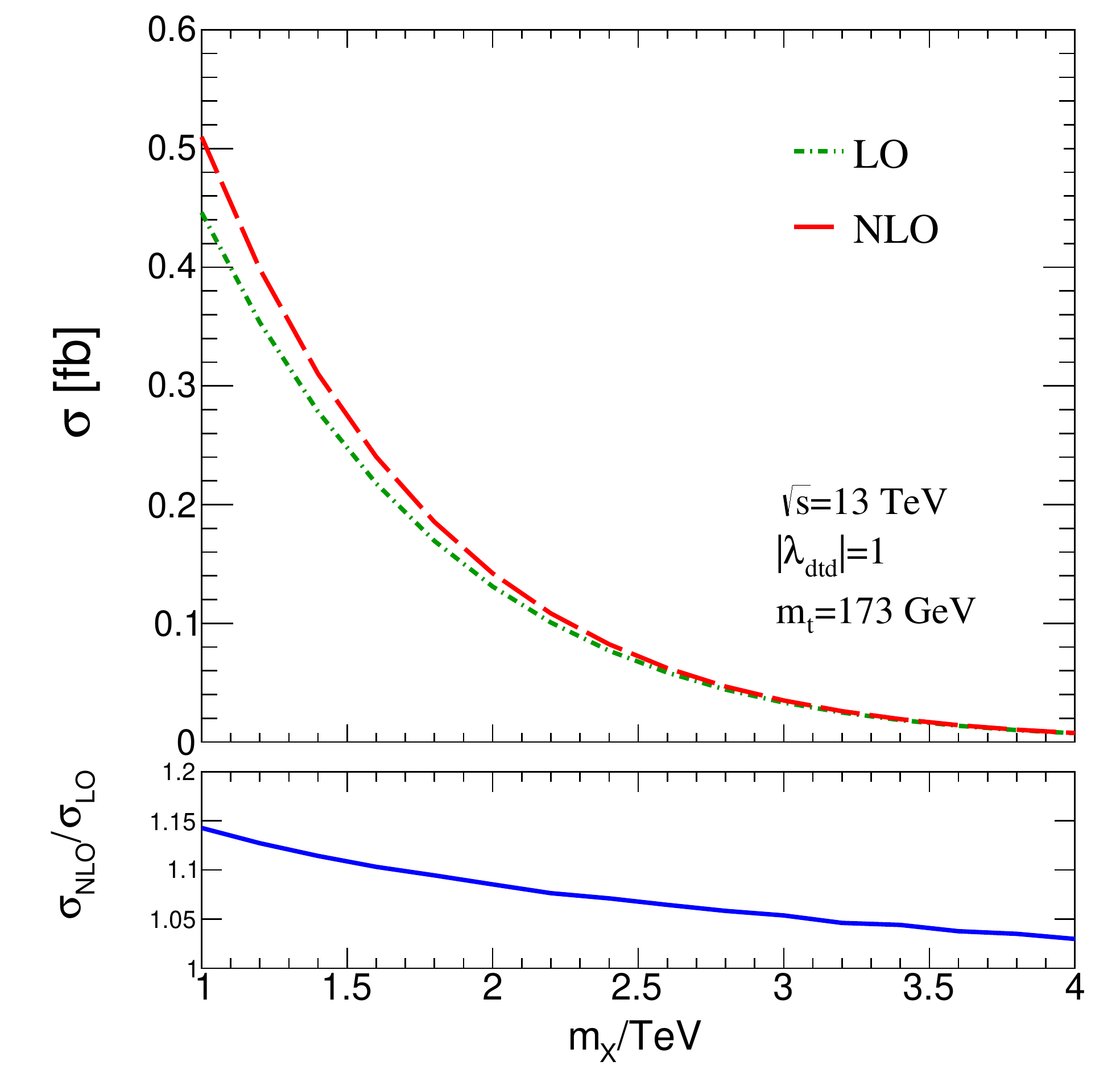}
			\caption{Cross section of mono-top process induced by ${\cal O}_{dtd}$ at the 13 TeV LHC.}\label{fig:monotopNLO}
		\end{center}
\end{figure}

The fixed order result for mono-top process is shown in Fig.~\ref{fig:monotopNLO} up to NLO level. The inclusive cross section decreases from 0.5 ${\rm fb}$ to 0.007~${\rm fb}$, and the $K$-factor also decreases from 1.14 to 1.03 as $m_X$ increases from 1 TeV to 4 TeV. Because the branch ratio of $t\to b+W(\to l\,\nu_l), (l=e, \mu)$ is about 20\%, we can estimate that the inclusive cross section of mono-top signal is less than 0.1 fb. This helps us to discuss the constraints on ${\cal O}_{dtd}$ in the following section.
	
\subsection{Mono-jet analysis}
The mono-jet signature has been studied in detail by current experiments~\cite{ATLAS:2017dnw}, where the analysis was performed with an integrated luminosity of 36.1~${\rm fb}^{-1}$ at the 13 TeV LHC. Here, the DM signals are simulated by $\texttt{Madgraph5}$ with parton shower.
Events are selected with $E_{\rm miss}^T >$  250 GeV, where a leading (highest-$p_T$) jet with $p_T>$ 250 GeV and $|\eta|< 2.4$ is required.
Most of the SM backgrounds are from  $Z(\to\nu\nu)$+jets processes. $W(\to \tau\nu)$+jets processes also give significant contributions. Top pair, diboson, multijet and single top processes give small contributions. In Ref.~\cite{ATLAS:2017dnw}, the SM background $Z$+jets and $W$+jets are normalized to next-to-next-leading order (NNLO) QCD and NLO electroweak predictions. Other backgrounds are simulated at NLO QCD level by using MC generators Powheg-Box and MadGraph5\_aMC@NLO\cite{Alwall:2014hca}. In Tab.~\ref{monojetbk} , we extract the mono-jet background in various signal regions from Ref.~\cite{ATLAS:2017dnw}.
	
\begin{figure}
		\begin{center}
			\includegraphics[scale=0.4]{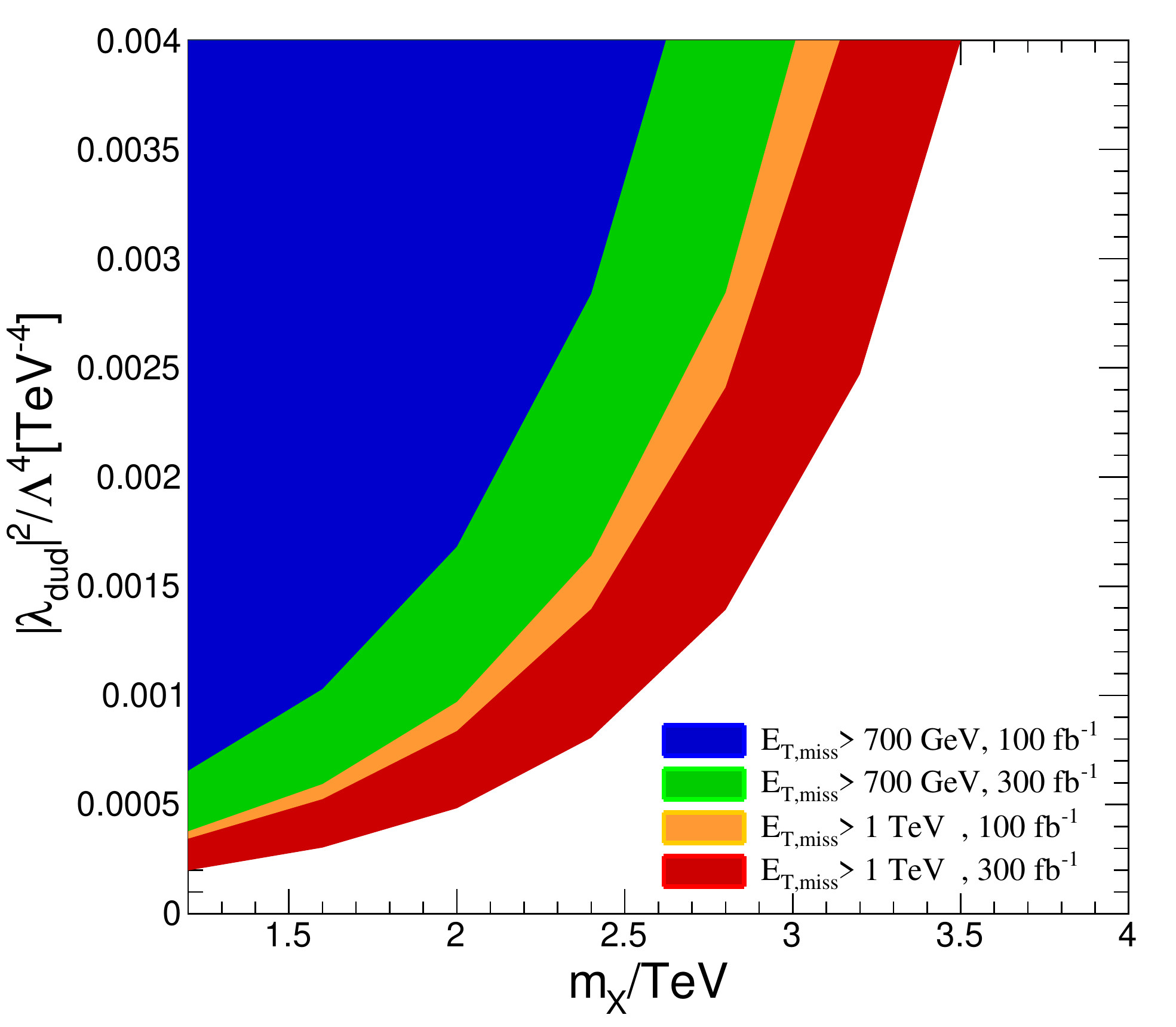}
			\includegraphics[scale=0.4]{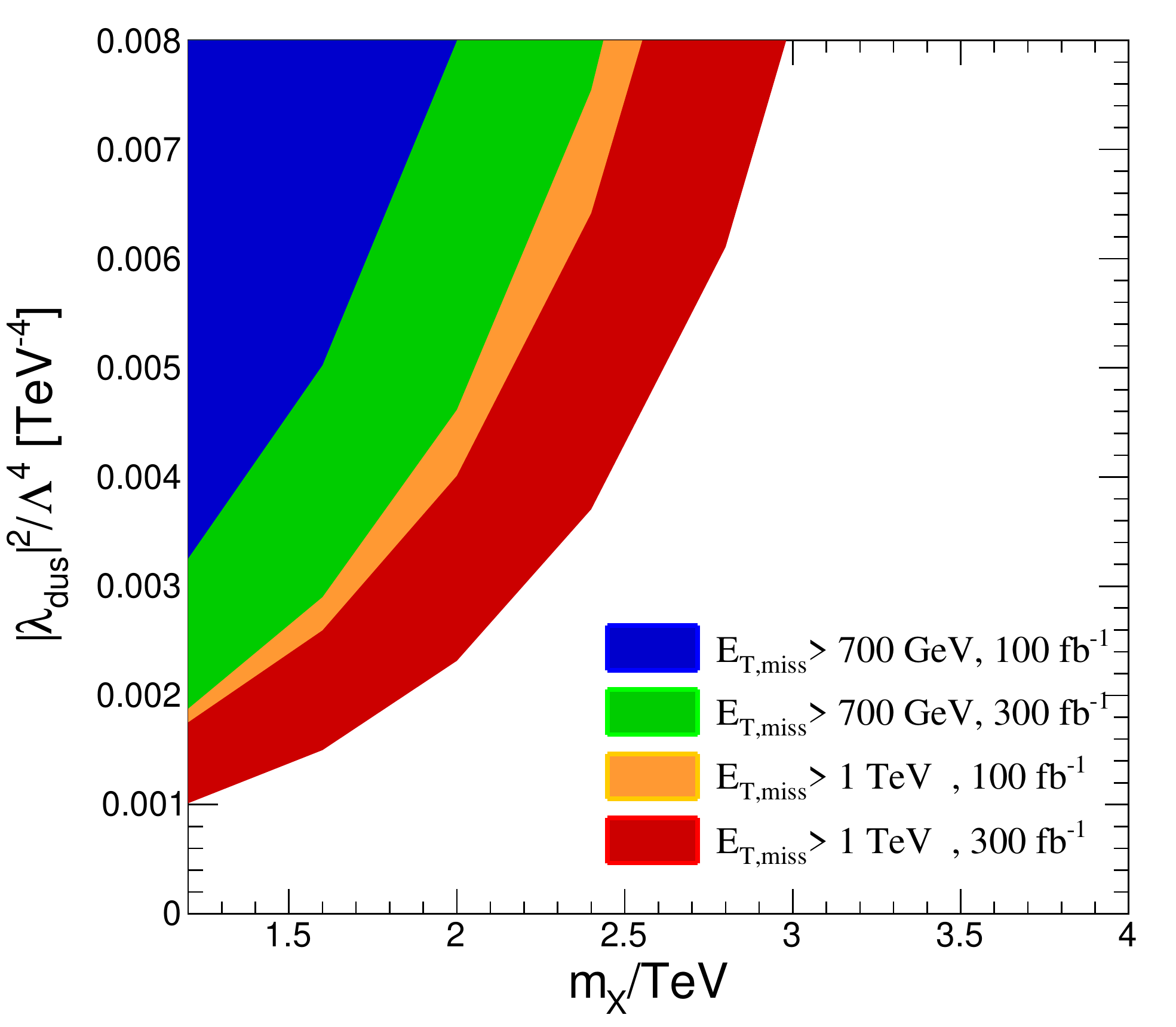}
			\caption{Constraints on coupling $\lambda_{ijk}$ and mass $m_X$ by mono-jet measurements at the 13 TeV LHC.}\label{fig:monojetcons}
		\end{center}
\end{figure}

In principle, if no signal is observed, the couplings $\lambda_{dud}$ and $\lambda_{dus(b)}$ cannot be too large.
Thus, in Fig.~\ref{fig:monojetcons}, we give 3$\sigma$ exclusion limits of the couplings against heavy Dirac fermion mass for integrated luminosity of $100\,{\rm fb}^{-1}$ and $300\,{\rm fb}^{-1}$ at the 13 TeV LHC by using the NLO theoretical predictions. The colored regions denote the parameters spaces that should be excluded if no signal is observed. The constraint for $\lambda_{dud}$ is stronger than $\lambda_{dud}$ because the former cross section is larger.
	
\begin{table}
		\begin{center}
			\begin{tabular}{ c| c c c c c}
				\hline
				\hline
				$\Eslash_T $[GeV] &~~~~ $>$ 250~~~~   &~~~~$>$ 300 ~~~~  & ~~~~$>$ 350 ~~~~  & ~~~~ $>$ 400 ~~~~ &~~~~ $>$ 500~~~~
				\\
				\hline
				~~~Background [${\rm fb}$] ~~~& 7077 &3997  & 2128 & 1150 & 378.9
				\\
				\hline
				\hline
				$\Eslash_T $[GeV] & $>$ 600   & $>$ 700   & $>$ 800   & $>$ 900  & $>$ 1000
				\\
				\hline
				Background [${\rm fb}$] & 141.2&58.78 &27.15&12.96&6.787
				\\
				\hline
				\hline
			\end{tabular}
		\end{center}
		\caption{\label{monojetbk} SM background predictions in the signal region for several inclusive $\Eslash_T$ selections.}
\end{table}
	
\subsection{Mono-top analysis}
Mono-top signals can be explored by using hadronic or semi-leptonic top decay modes~\cite{Andrea:2011ws,Wang:2011uxa}. For highly boosted mono-top production, we can take advantage of the jet substucture technique to perform top reconstruction and suppress the multi-jet background~\cite{CMS:2016flr,CMS:2017gin}. For un-boosted top, semi-leptonic can be used due to the clean leptonic signature. In this work, we do analysis with semi-leptonic top decay modes.  The background contributions are mainly from $W$+ jets, single top, top pair and gauge boson pair productions. Neutrino from $W$, $Z$  and top decay results in missing transverse energy. The dominant background is from $W$+jets process because of its huge cross section, so $b$-tagging must be performed to suppressed this background. Here, the backgrounds are generated with 0/1/2 jet parton level matching, based on the default $k_T$ -jet MLM scheme in MadGraph5\_aMC@NLO.
	
\begin{figure}
		\begin{center}
			\subfigure[$p_T^l$]{
				\label{fig:plcut}\includegraphics[scale=0.38]{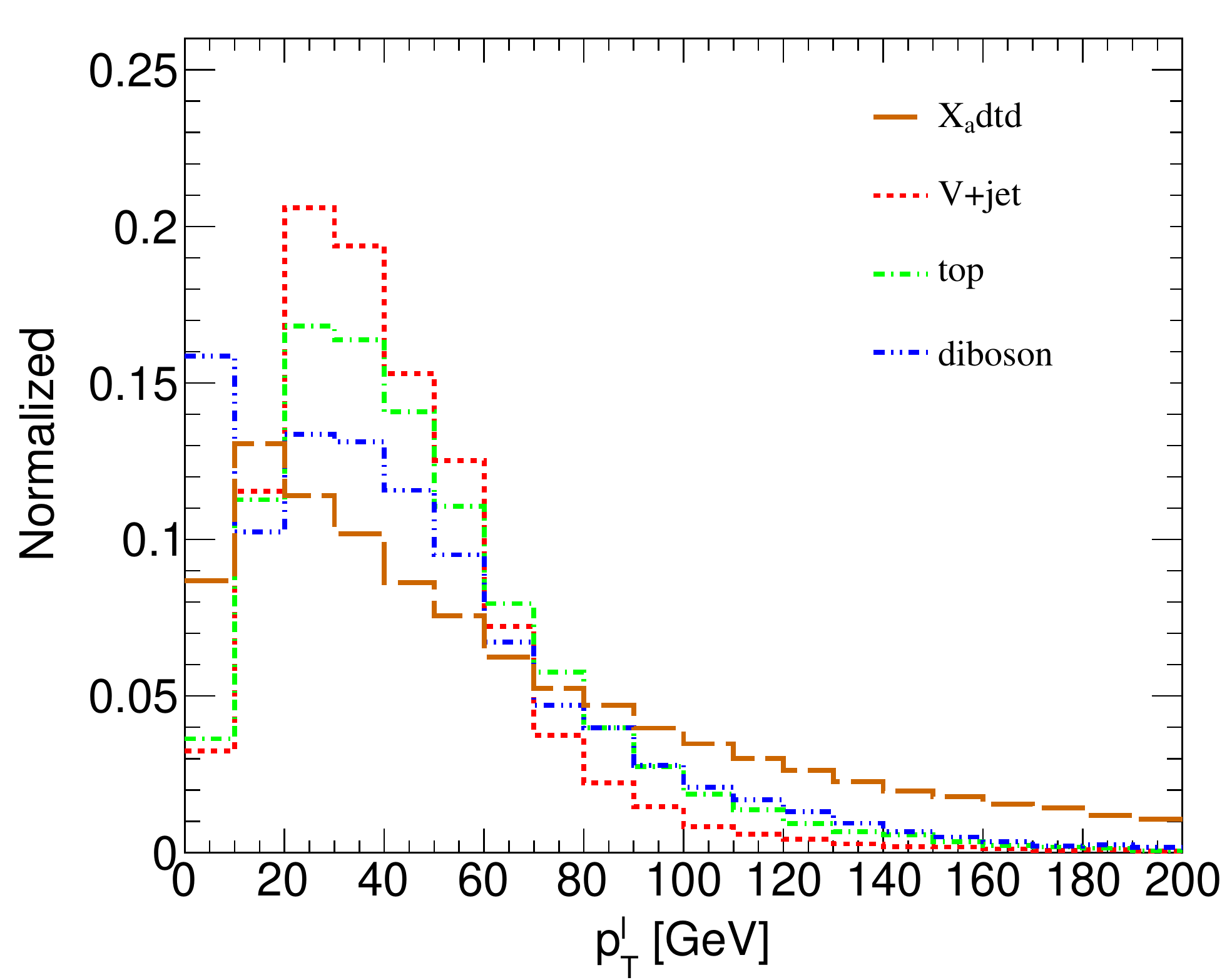}
			}
			\subfigure[$\Eslash_T$]{
				\label{fig:etmisscut}\includegraphics[scale=0.38]{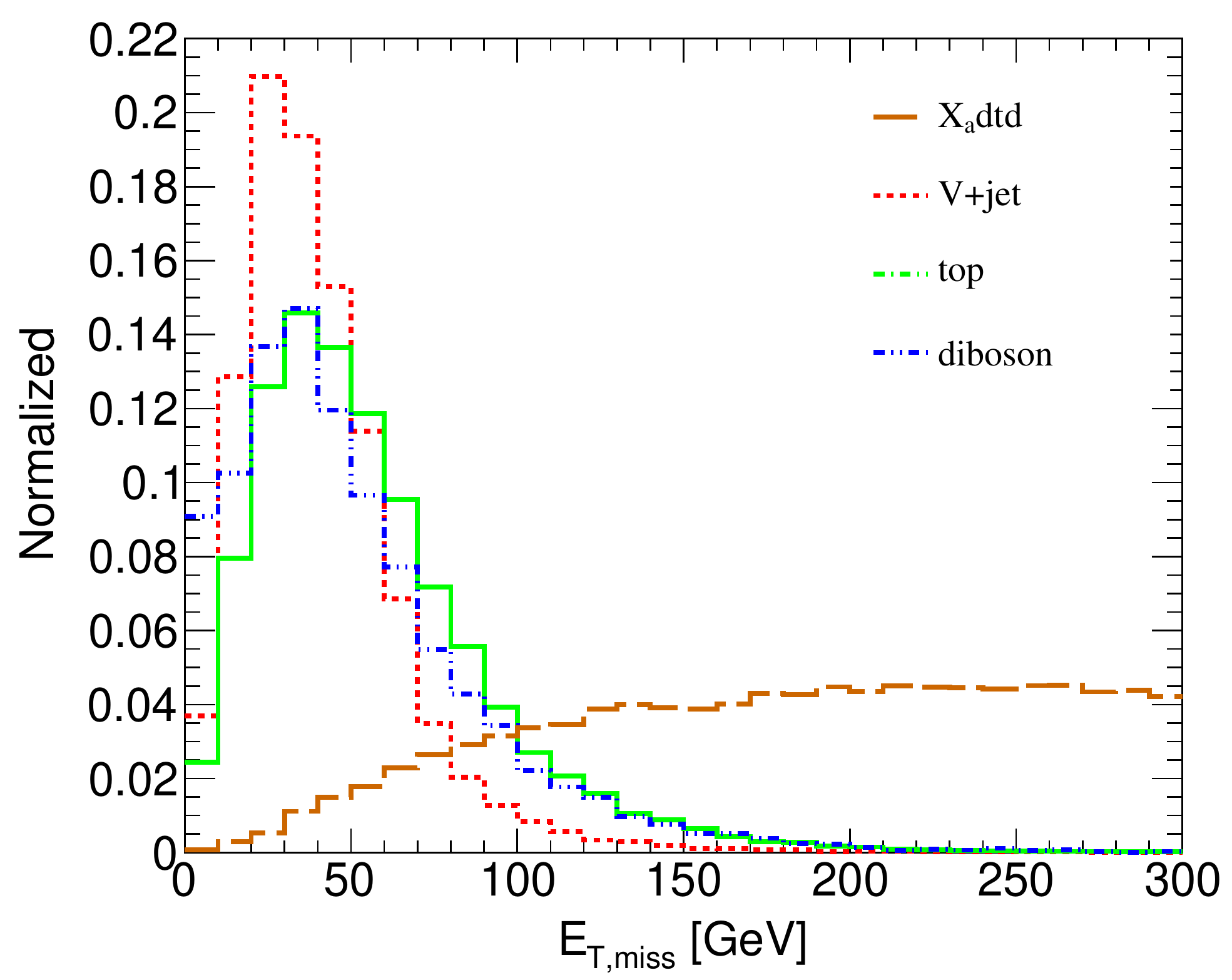}
			}\\
			\subfigure[$M_T$]{
				\label{fig:mtranscut}\includegraphics[scale=0.38]{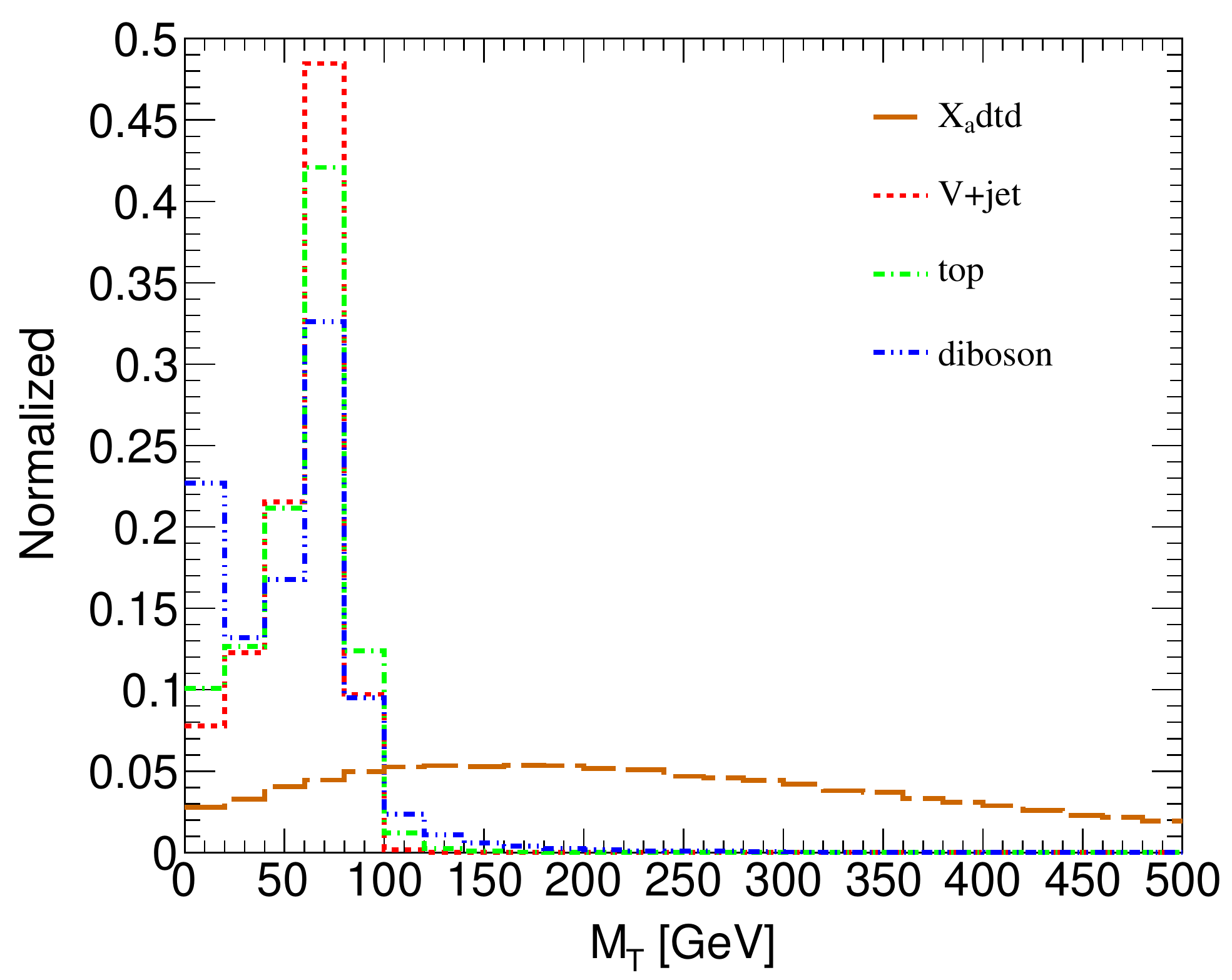}
			}
			\caption{Normalized spectra for signal and background in mono-top searching at the 13 TeV LHC. ``Top'' denotes the sum of background for top pair, single top and associated production of $tW$.}\label{fig:monotopcut}
		\end{center}
\end{figure}
	
Firstly, we introduce the basic cuts
\begin{gather}
	p_T^b>70\,{\rm GeV}\,, \quad \quad \quad |\eta^{b,l}|<2.4 \,\,\,.
\end{gather}
The selected leptons should be isolated, having  $\sum_i p_{T,i}$ less than 10\% of its transverse momentum within a cone of $\Delta R$ = 0.3 around it.
	
\begin{figure}
		\begin{center}
			\includegraphics[scale=0.40]{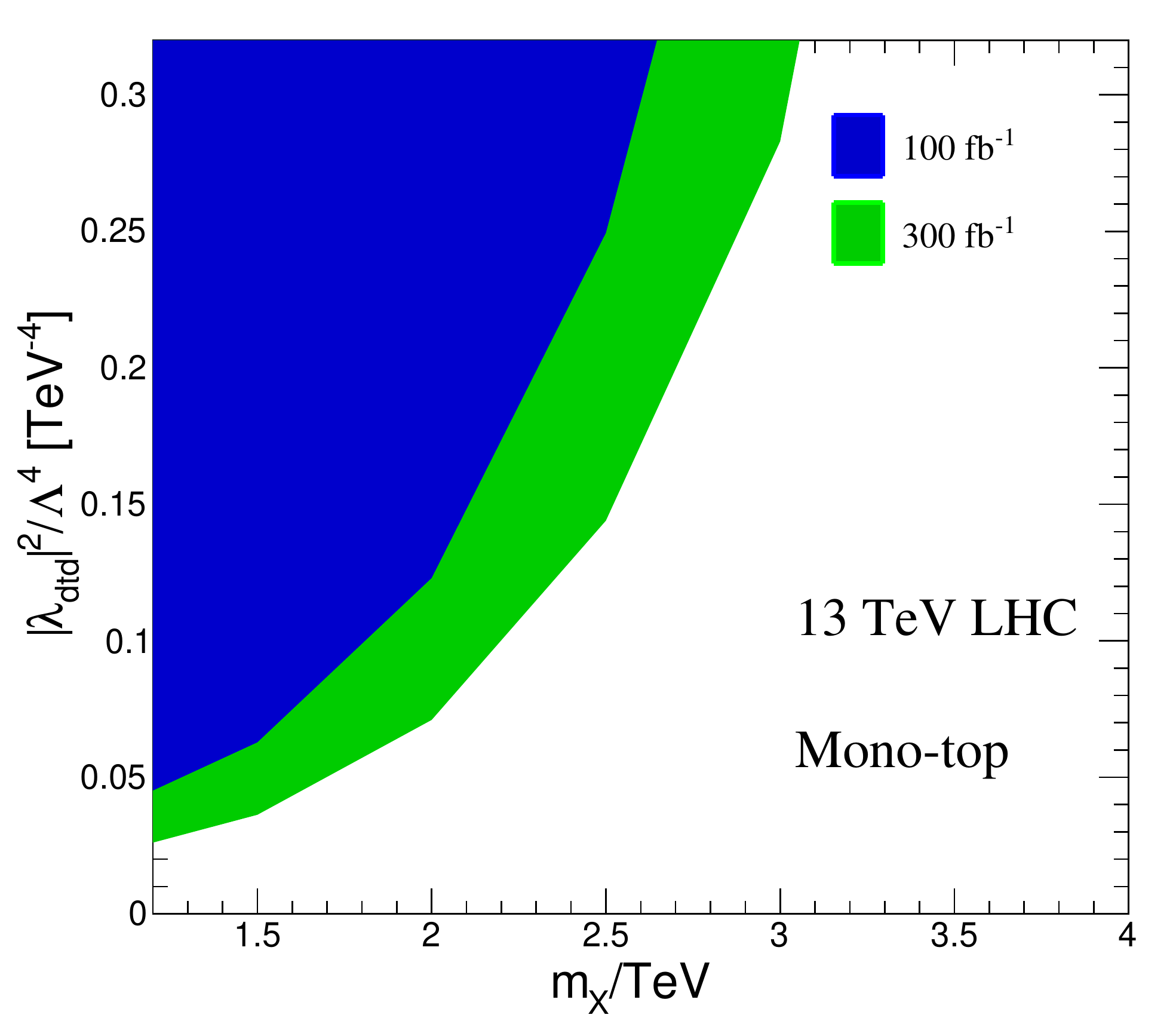}
			\caption{Normalized spectra for signal and background in mono-top searching at the 13 TeV LHC.}\label{fig:Xdtdcons}
		\end{center}
\end{figure}
	
The normalized transverse momentum distribution of lepton $e$ or $\mu$ in the semi-leptonic mode are shown in Fig.~\ref{fig:plcut}. There is no significant differences between various backgrounds and signals. So we choose a loose cut
\begin{gather}
	p_T^l>30\,{\rm GeV}\,.
\end{gather}
In addition, we veto extra lepton $e$ or $\mu$ with $p_T^l>30\,{\rm GeV}$ to suppress the background from $pp\to W(\to l\nu_l)+ W(\to l\nu_l) + j$, $pp\to W(\to l\nu_l)+Z(\to ll) + j$ and $pp\to W(\to l\nu_l)+t(\to b l\nu_l)$.
	
Fig.~\ref{fig:etmisscut} shows the normalized distribution of the missing transverse energy. The peak of background spectrum is around $m_W/2\approx 40\,{\rm GeV}$, because the (anti-)neutrino decay from $W$ boson takes half of its energy. While the missing energy for signal is significantly larger because two invisible particles are contained. Therefore, the missing transverse energy cut can be chosen as
\begin{gather}
\Eslash_T>100\,{\rm GeV}\,.
\end{gather}
	
In Fig.~\ref{fig:mtranscut}, we show the normalized transverse mass distribution for background and signal, which is defined with lepton and missing transverse momentum
\begin{gather}
	M_T=\sqrt{(\Eslash+E_T^l)^2-(\vec{\pslash_T}+\vec{p_T^l})^2}  \,\,\,.
\end{gather}
The spectra of various background have a peak around the $W$ boson mass $m_W\approx 80\,{\rm GeV}$, because the lepton and neutrino is from $W$ boson decay,  while the transverse mass spectrum of signal is smoothly distributed due to the fact that two invisible particles are contained. Therefore, the missing transverse energy cut can be performed as
\begin{gather}
M_T>100\,{\rm GeV}\,.
\end{gather}
	
In order to suppress the huge background of $W$+jets, $b$-tagging technique must be involved. The $b$-tagging efficiency is chosen as 70\%, and the light-jet-to-b miss-tagging probabilities are assumed of 1\%. Combining with the improved cuts of $p_T^l$, $\Eslash$ and $M_T$, the backgrounds of $V$+jets, top and diboson are 1.094 fb, 0.455 fb and 1.05 fb, respectively. The background from diboson is still significant because of the decay mode $W(\to l\nu_l)Z(\to \nu\nu)$.
Fig.~\ref{fig:Xdtdcons} gives the 3$\sigma$ exclusion limits of the couplings against heavy Dirac fermion mass for integrated luminosity of $100\,{\rm fb}^{-1}$ and $300\,{\rm fb}^{-1}$ at the 13 TeV LHC by using the NLO theoretical predictions. Comparing with $\lambda_{dud}$ and $\lambda_{dus}$,  the constraint is very weak, because the cross section of mono-top induced by ${\cal O}_{dtd}$ is much smaller.
	
\section{conclusion}\label{sec:sum}
We have studied the possibility of detecting the mechanism to solve the baryogenesis and DM with various large DM masses through strong first order phase transition and Q-balls.
The signals at GWs experiments and the LHC with QCD NLO accuracy have been discussed in detail.
We have found that the GWs could provide a realistic and complementary approach for testing the baryogensis and DM scenario.
Our results show that the phase transition process in the early universe may play an important role in solving the fundamental problems in particle cosmology. More systematical study on the phase transition physics in particle cosmology is left to our future study.	
\begin{acknowledgments}
We deeply appreciate Ze Long Liu's help on accomplishing the complicated QCD loop calculations.
F.P.H. thanks David J. Weir for his wonderful lectures on phase transition gravitational waves, and the useful discussions on Q-balls.
C.S.L is supported by the National Nature Science Foundation of China under Grant No. 11375013.
F.P.H. is supported by IBS under the project code IBS-R018-D1.
\end{acknowledgments}
\appendix
\section{Analytical results of NLO QCD corrections}\label{app1}
The virtual corrections contain both ultraviolet (UV) and infrared (IR) divergences, and the UV divergences
can be canceled by introducing counterterms. For the external fields, we fix all the renormalization constants using on-shell subtraction
\begin{equation}
\begin{aligned}
	\delta Z_{2}^q=&-\frac{\alpha_s}{3\pi}C_\epsilon\left(\frac{1}{\epsilon_{\rm UV}}-\frac{1}{\epsilon_{\rm IR}}\right)\,,\\
	\delta Z_{2}^t=&-\frac{\alpha_s}{3\pi}C_\epsilon \left(\frac{1}{\epsilon_{\rm UV}}+\frac{2}{\epsilon_{\rm IR}}+4+3\ln\left(\frac{\mu^2}{m_t^2}\right)\right)\,\,\,,
\end{aligned}
\end{equation}
with $C_\epsilon=\frac{(4\pi)^\epsilon}{\Gamma (1-\epsilon)}.$
For the coupling constants, we use the $\overline{\rm MS}$ scheme,
\begin{equation}
\begin{aligned}
	\delta Z_{\lambda_{dud}}=&\delta Z_{\lambda_{dus(b)}}=\delta Z_{\lambda_{dtd}}=-\frac{\alpha_s}{2\pi}C_\epsilon\frac{1}{\epsilon_{\rm UV}}\,.
\end{aligned}
\end{equation}
	
For the operator ${\cal O}^{dud}$, the UV renormalized one-loop virtual QCD correction can be expressed as
\begin{equation}\label{Xdudvirt}
\begin{aligned}
	i{\cal M}_{ud\to d X_a}^{\rm virt}=&i{\cal M}_{ud\to d X_a}^{\rm born}\times\frac{\alpha_s}{4\pi}C_\epsilon\Bigg\{-\frac{4}{\epsilon^2}-\frac{2}{3\epsilon}\left[2\ln\left(-\frac{\mu^2}{s}\right)+2\ln\left(-\frac{\mu^2}{t}\right)+2\ln\left(-\frac{\mu^2}{u}\right) + 9\right]\\
	& -\frac{2}{3}\Bigg[\ln\left(-\frac{\mu^2}{s}\right)^2+\ln\left(-\frac{\mu^2}{t}\right)^2+\ln\left(-\frac{\mu^2}{u}\right)^2+2\ln\left(-\frac{\mu^2}{t}\right)+4\ln\left(-\frac{\mu^2}{u}\right)\\
	& +\frac{2\left(m_X^2(2t-s)+s^2+st-2t^2\right)}{m_X^2(s+t)-(s-t)^2}\ln\left(-\frac{t}{s}\right)+14\Bigg]
	\Bigg\}
\end{aligned}
\end{equation}
For the operators ${\cal O}^{dus}$ and ${\cal O}^{dub}$, the UV renormalized one-loop virtual QCD correction can be expressed as
\begin{equation}\label{Xdusvirt}
\begin{aligned}
	&i{\cal M}_{us\to d X_a}^{\rm virt}=i{\cal M}_{us\to d X_a}^{\rm born}\times\frac{\alpha_s}{4\pi}C_\epsilon\Bigg\{-\frac{4}{\epsilon^2}-\frac{2}{3\epsilon}\left[2\ln\left(-\frac{\mu^2}{s}\right)+2\ln\left(-\frac{\mu^2}{t}\right)+2\ln\left(-\frac{\mu^2}{u}\right) + 9\right]\\
	& -\frac{2}{3}\Bigg[\ln\left(-\frac{\mu^2}{s}\right)^2+\ln\left(-\frac{\mu^2}{t}\right)^2+\ln\left(-\frac{\mu^2}{u}\right)^2+6\ln\left(-\frac{\mu^2}{t}\right)-\frac{2(m_X^2-s+t)}{s-m_X^2}\ln\left(\frac{t}{u}\right)+14\Bigg]
	\Bigg\}
\end{aligned}
\end{equation}
For the operator ${\cal O}^{dtd}$, the UV renormalized one-loop virtual QCD correction can be expressed as
\begin{equation}
	\begin{aligned}
	&i{\cal M}_{dd\to t X_a}^{\rm virt}=i{\cal M}_{dd\to t X_a}^{\rm born}\times\frac{\alpha_s}{4\pi}C_\epsilon\Bigg\{\frac{8}{-3\epsilon^2}-\frac{4}{3\epsilon}\Bigg[\log\left(-\frac{\mu^2}{s}\right)+\log\left(\frac{\mu^2}{m_t^2}\right)+\log\left(\frac{m_t^2}{m_t^2-u}\right)\\
	&\qquad +\log\left(\frac{m_t^2}{m_t^2-t}\right)\Bigg]-\frac{2}{3} \log ^2\left(\frac{\mu ^2}{m_t^2}\right)-\frac{2}{3} \log ^2\left(-\frac{\mu
		^2}{s}\right)-\frac{8}{3} \log \left(-\frac{\mu ^2}{s}\right)\\
	&\qquad-\frac{2}{3} \log \left(\frac{\mu ^2}{m_t^2}\right) \left[2 \log \left(-\frac{m_t^2}{m_X^2-s-t}\right)+2 \log
	\left(\frac{m_t^2}{m_t^2-t}\right)+1\right]\\
	&\qquad-\frac{2}{3} \log ^2\left(-\frac{m_t^2}{m_X^2-s-t}\right)-\frac{4}{3} \log^2\left(\frac{m_t^2}{m_t^2-t}\right)\\
	&\qquad-\frac{4}{3}\log \left(\frac{m_t^2}{m_t^2-t}\right) \frac{\left(2 m_t^4
		\left(m_X^2-t\right)+m_t^2 s \left(m_X^2-2 t\right)+t \left(-2 m_X^2 (s+t)+2 s^2+5 s t+2 t^2\right)\right)}{ t
		\left(m_t^2 \left(4 m_X^2-s-4 t\right)-m_X^2 (s+4 t)+(s+2 t)^2\right)}\\
	&\qquad-\frac{4}{3}\log \left(-\frac{m_t^2}{m_X^2-s-t}\right)\Bigg[\frac{m_t^4 \left(4 m_X^2-s-4
		t\right)+m_t^2 \left(2 m_X^4-m_X^2 (s+8 t)+2 t (2 s+3 t)\right)}{ \left(m_t^2+m_X^2-s-t\right)
		\left(m_t^2 \left(4 m_X^2-s-4 t\right)-m_X^2 (s+4 t)+(s+2 t)^2\right)}\\
	&\qquad+\frac{ m_X^4 (s-2 t)-2 m_X^2 \left(s^2-2
		t^2\right)+s^3+2 s^2 t-s t^2-2 t^3}{ \left(m_t^2+m_X^2-s-t\right)
		\left(m_t^2 \left(4 m_X^2-s-4 t\right)-m_X^2 (s+4 t)+(s+2 t)^2\right)}\Bigg]\\
	& \qquad-\frac{2}{9} \left[-6
	\text{Li}_2\left(\frac{m_t^2+m_X^2-s-t}{m_X^2-s-t}\right)-6 \text{Li}_2\left(-\frac{t}{m_t^2-t}\right)+\pi
	^2+33\right]\Bigg\}  \nonumber \,\,\,.
\end{aligned}
\end{equation}
	
\bibliographystyle{apsrev}
\bibliography{reference}
\end{document}